\documentclass[12pt]{article}

\usepackage[portuges]{babel}
\usepackage{graphicx, xcolor, amsmath, amssymb, amsthm, amsfonts}
\usepackage[hmarginratio=6:6,top=30mm,columnsep=12mm,textwidth=170mm]{geometry}
\usepackage{multicol} 
\usepackage[hang,small,labelfont=bf,up,textfont=it,up]{caption} 
\usepackage{booktabs} 
\usepackage{float} 
\usepackage{hyperref} 
\usepackage{paralist} 
\usepackage{titlesec} 
\usepackage{fancyhdr} 
\title{\vspace{-13mm}\fontsize{16pt}{11pt}\selectfont \textbf{Solu\c{c}\~oes anal\'iticas da equa\c{c}\~ao de Schr\"odinger para um potencial hiperb\'olico deform\'avel\\ \vspace{2mm}
{\normalsize (Analytical solutions for the Schr\"odinger equation subjected to a deformable hyperbolic potential)}}} 
\vspace{-3mm}
\author{
\large {C. J. M. Fernandes}\thanks{Email:carlajamile\_melo@hotmail.com}, M. S. Cunha\thanks{Email:marcony.cunha@uece.br} 
 \\ \vspace{-1mm}
\small  Grupo de F\'isica Te\'orica - GFT, Centro de Ciência e Tecnologia - CCT\\ \small Universidade Estadual do Cear\'a - UECE\\ \small Av. Dr. Silas Munguba, 1700, CEP 60914-903, Fortaleza - CE.
}
\date{}
\begin{document}
\maketitle
\thispagestyle{fancy} 
\newcommand{\beq}{\begin{equation}}
\newcommand{\eeq}{\end{equation}}
\newcommand{\bea}{\begin{eqnarray}}
\newcommand{\eea}{\end{eqnarray}}
\noindent{\footnotesize{\textbf{Resumo:} Neste trabalho discutimos detalhadamente as conhecidas solu\c{c}\~oes da equa\c{c}\~ao de Schr\"odinger estacion\'aria sujeita a um potencial hiperb\'olico deform\'avel exatamente sol\'uvel $V(x) = \frac{V_0}{2}(1+\tanh(\delta x))$. Encontramos as solu\c{c}\~oes anal\'iticas em termos das fun\c{c}\~oes hipergeom\'etricas de Gauss para os estados de espalhamento com energia maior que o m\'aximo do potencial. Discutimos tamb\'em o caso para energia menor que o m\'aximo e as semelhan\c{c}as e diferen\c{c}as com o potencial degrau abrupto para ambos os casos. Ilustramos graficamente as situa\c{c}\~oes f\'isicas relevantes para o problema.}}\\
{\footnotesize{\textbf{Palavras-chave:} Equa\c{c}\~ao de Schr\"odinger,  potencial hiperb\'olico tangente, fun\c{c}\~ao hipergeom\'etrica.}}
 \\[0.5cm]
\noindent{\footnotesize{\textbf{Abstract:} In this work we discuss in detail the known solutions of the stationary Schr\"odinger equation subject to a deformable hyperbolic tangent potential exactly soluble $ V(x) = \frac {V_0} {2} (1+ \tanh (\delta x)) $. We find the analytical solutions in terms of Gauss hypergeometric functions for the scattering states with energy greater than the maximum value of the potential. We also discussed the case for the energy lower than the maximum and the similarities and differences with the abrupt step potential in both cases. We graphically illustrate the relevant physical situations to the problem}.} \\
{\footnotesize{\textbf{Keyworks:} Schr\"odinger equation, hyperbolic tangent potential, hypegeometric functions.}}
\section{Introdu\c{c}\~ao}
\noindent\hspace{1.5cm}Lugar comum dizer que a Mec\^anica Qu\^antica (MQ) desempenha papel central na F\'isica, especialmente no que tange a seu grande leque de aplica\c{c}\~oes. Por\'em, no que concerne ao ensino de F\'isica, em particular \`a MQ, ainda caminhamos a passos lentos para sua moderniza\c{c}\~ao e amplia\c{c}\~ao, n\~ao s\'o porque as iniciativas foram muito centralizadas em determinadas regi\~oes do pa\'is, mas tamb\'em porque foram poucas. Vale ressaltar, entretanto, a iniciativa recente da Sociedade Brasileira de F\'isica em descentralizar a forma\c{c}\~ao de professores atrav\'es dos Mestrados Profissionais \cite{MNPEF}, o que veio somar aos j\'a consolidadas mestrados e doutorados acad\^emicos na \'area de Ensino de F\'isica.  Ainda assim, professores com forma\c{c}\~ao em Ensino de F\'isica ainda s\~ao raros em nossos cursos de F\'isica, mesmo naqueles que possuem licenciaturas. Laborat\'orios did\'aticos sem experimentos de F\'isca Moderna ainda s\~ao a regra em nossos cursos.	

\noindent\hspace{1.5cm}Passando um pouco ao largo desta quest\~ao hist\'orica do ensino de F\'isca no Brasil, neste trabalho pretendemos resgatar as solu\c{c}\~oes da equa\c{c}\~ao de Schr\"odinger independente do tempo para o potencial hiperb\'olico tangente, deform\'avel. Este potencial deform\'avel se aproxima do potencial degrau \cite{RIBEIRO} no limite de grandes valores do que chamamos aqui de par\^ametro de deforma\c{c}\~ao. Este problema foi descrito, at\'e onde sabemos, pela primeira vez em 1947, em uma publica\c{c}\~ao did\'atica em alem\~ao, e depois republicado em ingl\^es em 1971, com reedi\c{c}\~oes em 1991 e em 1999, por Siegfried Fl\"ugge \cite{FLUGGE}. 

\noindent\hspace{1.5cm} Potenciais hiperb\'olicos est\~ao relacionados, em grande parte, a problemas em f\'isica at\^omica e molecular. Em particular, podem ser utilizados para modelar transi\c{c}\~oes de dopagens abruptas ou graduais em jun\c{c}\~oes semicondutoras, ou ainda para modelar estruturas como nanofios com raio ou composi\c{c}\~ao (ou ambos) vari\'aveis \cite{WILLATZEN}. Existem vers\~oes hiperb\'olicas dos potenciais moleculares de Scarf \cite{SCARF}, Rosen-Morse \cite{ROSEN} e Manning-Rosen \cite{MANNING}, que s\~ao tamb\'em de interesse na modelagem de vibra\c{c}\~oes e de for\c{c}as moleculares \cite{MILOSLAV, YESILTAS, DUTRA1,GAO,WEN,CORREA,BHARALI}. 

\noindent\hspace{1.5cm} Modelos que envolvem potenciais hiperb\'olicos s\~ao tamb\'em muito utilizados em uma vasta gama de trabalhos encontrados em literatura especializada, nos mais variados contextos \cite{CLARA,DONG,DONG2, ARDA,ILDE,ISHKHANYAN}, inclusive incorporando diferentes t\'ecnicas para o estudo das solu\c{c}\~oes da equa\c{c}\~ao de Schr\"odinger \cite{CUNHA, CHRISTIANSEN, CHRISTIANSEN2,ABBAS,BAGCHI,DUTRA}. Obviamente, o estudo de solu\c{c}\~oes anal\'iticas tem uma import\^ancia fundamental no entendimento conceitual da f\'isica desses modelos e de suas aplica\c{c}\~oes.

\noindent\hspace{1.5cm} Encontra-se assim dividido este trabalho. Na pr\'oxima se\c{c}\~ao, estudamos os limites assint\'oticos da equa\c{c}\~ao de Schr\"odinger com o potencial hiperb\'olico, considerando ondas incidindo pela esquerda. Na se\c{c}\~ao seguinte, por meio de mudan\c{c}as nas vari\'aveis dependente e independente, mostraremos como obter as solu\c{c}\~oes anal\'iticas  reescrevendo a equa\c{c}\~ao de Schr\"odinger na forma da equa\c{c}\~ao hipergeom\'etrica de Gauss. A seguir, calculamos os coeficientes de reflex\~ao e transmiss\~ao para a energia maior que a barreira de potencial e reobtemos a express\~ao do potencial degrau como caso limite. Na se\c{c}\~ao seguinte, calculamos o coeficiente de reflex\~ao para energia menor que a barreira e mostramos que ele \'e identicamente igual \`a unidade, ou seja, que apesar da part\'icula penetrar na barreira de potencial, a reflex\~ao \'e total. Calculamos ainda as constantes que aparecem no problema, para este caso, aplicando as condic\~oes de continuidade da fun\c{c}\~ao de onda e de sua derivada na origem. Mais uma vez, reobtemos uma expressão para as constantes do potencial degrau quando o par\^amtero de deforma\c{c}\~ao cresce ($\delta \gg 1$). Por \'ultimo apresentamos nossas considera\c{c}\~oes finais.

\section{Potencial hiperb\'olico} \label{sec_pot}

\begin{figure}[h]
\center
{\includegraphics[width=8cm,height=8cm]{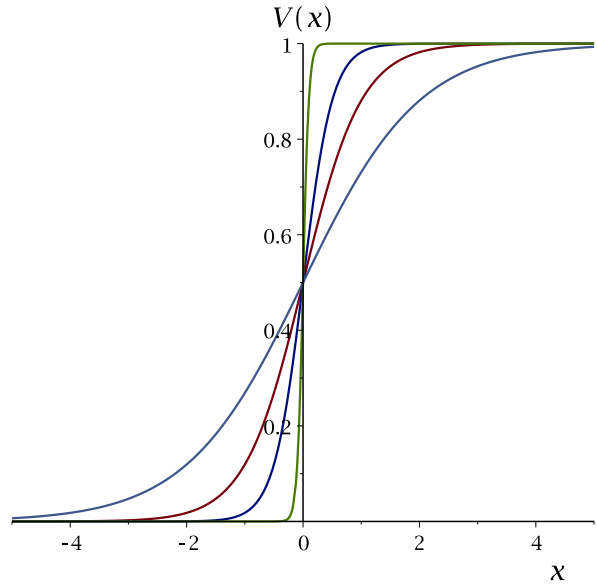}}\hspace{1cm}
\caption{\label{fig_pot} Potencial $V(x)$ para diversos valores do par\^ametro de deforma\c{c}\~ao $\delta$, a saber, $\delta=1/2,~1,~2,~10$ e $V_0=1$. Quanto maior $\delta$, mais o potencial se aproxima do potencial degrau.}
\end{figure}

\noindent\hspace{1.5cm} Neste trabalho, nosso objetivo \'e discutir em detalhes as (n\~ao t\~ao) conhecidas, nem discutidas nos livros did\'aticos, solu\c{c}\~oes anal\'iticas da equa\c{c}\~ao de Schr\"odinger para um potencial barreira tipo tangente hiperb\'olica deform\'avel, tanto para energias acima da barreira de potencial, quanto para os estados com energia abaixo do m\'aximo do potencial, para diversos valores dos par\^ametro de deforma\c{c}\~ao. Para isso, consideremos ent\~ao o potencial hiperb\'olico tipo degrau suave
\beq V(x) = \frac{V_0}{2}\left[1+\tanh(\delta x)\right], \label{pot}\eeq
%
onde $\delta$ \'e o par\^ametro que controla a deforma\c{c}\~ao do potencial, ou seja, o quanto o potencial $V(x)$ se aproxima ou se afasta de uma fun\c{c}\~ao degrau (quanto maior o valor de $\delta$ mais esse potencial se aproxima de um degrau abrupto) [veja Fig. (\ref{fig_pot}) abaixo].\\
\\
A equa\c{c}\~ao de Schr\"odinger estacion\'aria para este potencial \'e dada por
\beq
-\frac{\hbar^2}{2m}\frac{d^2\psi(x)}{dx^2} +\frac{V_0}{2}\left[1+\tanh(\delta x)\right] \psi(x) =E\psi(x). \label{sch}
\eeq
\noindent\hspace{1.5cm}\'E \'util, antes de procurar por solu\c{c}\~oes anal\'iticas da equa\c{c}\~ao, analisar seus limites assint\'oticos. Assim, tomando o limite $x\rightarrow-\infty$, temos aproximadamente uma equa\c{c}\~ao de part\'icula livre
\beq
-\frac{\hbar^2}{2m}\frac{d^2\psi(x)}{dx^2}\approx E \psi(x). \label{sch_assim}
\eeq
cuja solu\c{c}\~ao \'e
\beq \label{assim_L}
\psi\approx A e^{ik x}+ B e^{-ik x}
\eeq
onde $k^2=2mE/\hbar^2$.

\noindent\hspace{1.5cm}O comportamento em $x\rightarrow\infty$ \'e semelhante, com equa\c{c}\~ao assint\'otica dada por
\beq \label{sch_approx}
-\frac{\hbar^2}{2m}\frac{d^2\psi(x)}{dx^2}\approx (E-V_0) \psi(x). 
\eeq
A solu\c{c}\~ao da equa\c{c}\~ao acima, para $E>V_0$ e para ondas incidindo pela esquerda,  \'e 
\beq \label{assim_R}
\psi\approx C e^{i\ell x}\!\!\!\!\!\!,
\eeq
onde $\ell^2=2m(E-V_0)/\hbar^2$.\\

\noindent\hspace{1.5cm}Para $E<V_0$, a solu\c{c}\~ao fisicamente aceit\'avel deve se atenuar para $x\rightarrow \infty$. Portanto, temos
\beq \label{pb}
\psi(x)\approx D\, e^{-\kappa x}
\eeq
onde $\kappa^2=2m (V_0-E)/h^2$.\\

\section{Solu\c{c}\~oes anal\'iticas: caso $E>V_0$}

\noindent\hspace{1.5cm}Para encontrar as solu\c{c}\~oes anal\'iticas da equa\c{c}\~ao de Schr\"odinger, Eq. (\ref{sch}), para o caso da energia maior que a barreira de potencial, podemos usar a seguinte mudan\c{c}a de vari\'aveis
\beq
y=-e^{- 2 \delta x}, \label{transy}
\eeq
e assim reescrever a equa\c{c}\~ao de Schr\"odinger como 
\beq \label{eqSy}
y^2\psi''(y)+y\psi'(y)+ \frac{2m}{\hbar^2}\frac{1}{4\delta^2}\left(E - \frac{V_0}{1-y}\right)\psi(y) = 0.
\eeq
O passo seguinte \'e realizar uma transforma\c{c}\~ao na vari\'avel dependente, a saber,
\beq
\psi(y)=y^\alpha(1-y)^\beta F(y), \label{transf} \eeq
na qual $F(y)$ \'e uma fun\c{c}\~ao desconhecida. Substituindo a transforma\c{c}\~ao acima na Eq. (\ref{eqSy}), temos
\bea \label{geral}%
y\, (1-y)\,F''(y)\!&+&\!\left[1+2\alpha\!-\!(1+2\beta+2\alpha)y\right]F'(y)\nonumber\\
&+&\!\!\!\left[\left(\alpha^2\!+\!\frac{\mathcal{E}}{4\delta^2}\!-\!\frac{\mathcal{V}_0}{4\delta^2}\right)\frac{1-y}{y}\!+\!\frac{\beta(\beta\!-\!1)y}{1\!-\!y}- (2\alpha+1)\beta\!-\!\frac{\mathcal{V}_0}{4\delta^2}\right]\!\!F(y)=0,
\eea 
onde definimos $\mathcal{E}=2mE/\hbar^2$ e $\mathcal{V}_0=2mV_0/\hbar^2$. Sem perda de generalidade, podemos assumir
\begin{subequations}
\bea
\beta=1 \\
\alpha^2+\mathcal{E}/4\delta^2-\mathcal{V}_0/4\delta^2=0.\label{eq_alpha}
\eea
\end{subequations}
tal que $\alpha=\pm\, (i/2\delta) \sqrt{\mathcal{E}-\mathcal{V}_0}$. Potanto, a Eq. (\ref{geral}) acima assume ent\~ao a forma de uma equa\c{c}\~ao hipergeom\'etrica de Gauss,
\bea
y(1-y)\,F''(y) + \left[1+2\alpha-(3+2\alpha)\,y\right]F'(y)
-\left(1+2\alpha+\mathcal{V}_0/4\delta^2\right)\!F(y)=0 \label{gauss}
\eea
cuja forma geral padr\~ao \'e 
\beq \label{hypergeom}
z(1-z)\,F''(z) + [c-(1+a+b)z ] F'(z) - a b F(z)=0,
\eeq
onde $a$ $b$ e $c$ s\~ao constantes. A solu\c{c}\~ao geral desta equa\c{c}\~ao \'e dada em termos das fun\c{c}\~oes hipergeom\'etricas de Gauss \cite{ABRAMOWITZ}, a saber, 
\bea
F(z)= {}_2F_1 (a,b,c;\,z)+z^{1-c}{}_2F_1(a-c+1,b-c+1, 2-c;\,z).
\eea
Comparando as Eqs. (\ref{gauss}) e (\ref{hypergeom}) acima, vemos facilmente que
\begin{subequations}
\bea
c = 1+2\alpha\\
a+b=2+2\alpha\\
a\, b = 1+2\alpha+\mathcal{V}_0/4\delta^2
\eea
\end{subequations}
Resolvendo o sistema para $a$ e $b$, e utilizando a solu\c{c}\~ao da Eq (\ref{eq_alpha}), obtemos 
\begin{subequations}
\bea
a=1+\alpha\pm i \frac{\sqrt{\mathcal{E}}}{2\delta}\label{eqa}\\
b=1+\alpha \mp i\frac{\sqrt{\mathcal{E}}}{2\delta}\label{eqb}
\eea
\end{subequations}
\noindent\hspace{1.5cm}Escolhendo arbitrariamente, sem perda de generalidade, $\alpha=-i\nu$, e o sinal superior nas equa\c{c}\~oes acima, ficamos com (o leitor interessado pode verificar que a escolha $\alpha=i\nu$ produz exatamente as mesmas solu\c{c}\~oes)
\begin{subequations}
\bea
a &\!\!=\!\!& 1+i(\mu-\nu)\\
b &\!\!=\!\!& 1-i(\mu+\nu)\\
c &\!\!=\!\!& 1-i2\nu
\eea
\end{subequations}
onde
\bea
\mu=\frac{\sqrt{\mathcal{E}}}{2\delta}; ~~~~ \nu=\frac{\sqrt{\mathcal{E}-\mathcal{V}_0}}{2\delta}\label{munu}
\eea 
Podemos ent\~ao escrever a solu\c{c}\~ao geral da Eq. (\ref{gauss}) como
\bea
 F(y)= C_1~{}_2F_1\Big(1+i(\mu-\nu),1-i(\mu+\nu),1-2i\nu; y\Big)\hspace{5cm} \nonumber\\
+ C_2~y^{2i\nu}{}_2F_1 \Big(1+i(\mu+\nu), 1-i(\mu-\nu), 1+2i\nu; y\Big).
\eea
Portanto, substituindo a equa\c{c}\~ao acima na Eq. (\ref{transf}) temos, %
\bea \label{eq_psi}
\psi (y) = C_1y^{-i\nu} (1-y)\,_2F_1\Big(1+i(\mu-\nu),1-i(\mu+\nu),1-2i\nu;y\Big)\hspace{3cm}\nonumber\\
+ C_2\, y^{i\nu}(1-y)
\,{}_2F_1 \Big(1+i(\mu+\nu),1-i(\mu-\nu),1+2i\nu;y\Big).
\eea
Em termos da vari\'avel $x$, a fun\c{c}\~ao de onda fica
\bea \label{eq_psix}
\psi (x) = C_1(-1)^{-i\nu} e^{i\nu2\delta x} (1+e^{-2\delta x})\,_2F_1\Big(1+i(\mu-\nu),1-i(\mu+\nu),1-2i\nu;-e^{-2\delta x}\Big)\nonumber\\
+ C_2(-1)^{i\nu}\, e^{-i\nu 2\delta x}(1+e^{-2\delta x})
\,{}_2F_1 \Big(1+i(\mu+\nu),1-i(\mu-\nu),1+2i\nu;-e^{-2\delta x}\Big).
\eea 

\noindent\hspace{1.5cm}De acordo com a solu\c{c}\~ao assint\'otica em $x\rightarrow \infty$, Eq. (\ref{assim_R}), \'e imediata a escolha da primeira solu\c{c}\~ao da equa\c{c}\~ao acima como a onda transmitida a partir de uma onda incidindo pela esquerda, uma vez que a segunda solu\c{c}\~ao refere-se a uma onda refletida vindo da direita. Portanto, para a onda transmitida \`a direita da origem, temos
\beq \label{psi_R}
\Psi_T=D\,\psi_{trans}
\eeq
onde 
\bea
\psi_{trans} (x) = e^{i\nu2\delta x}(1+e^{-2\delta x})\,{}_2F_1 \Big(1+i(\mu-\nu),1-i(\mu+\nu),1-2i\nu;-e^{-2\delta x}\Big). \label{eq_T}
\eea
e $D=(-1)^{-i\nu}C_1$.\\

\begin{figure}[ht] 
\center
{\includegraphics[width=6.8cm,height=6.8cm]{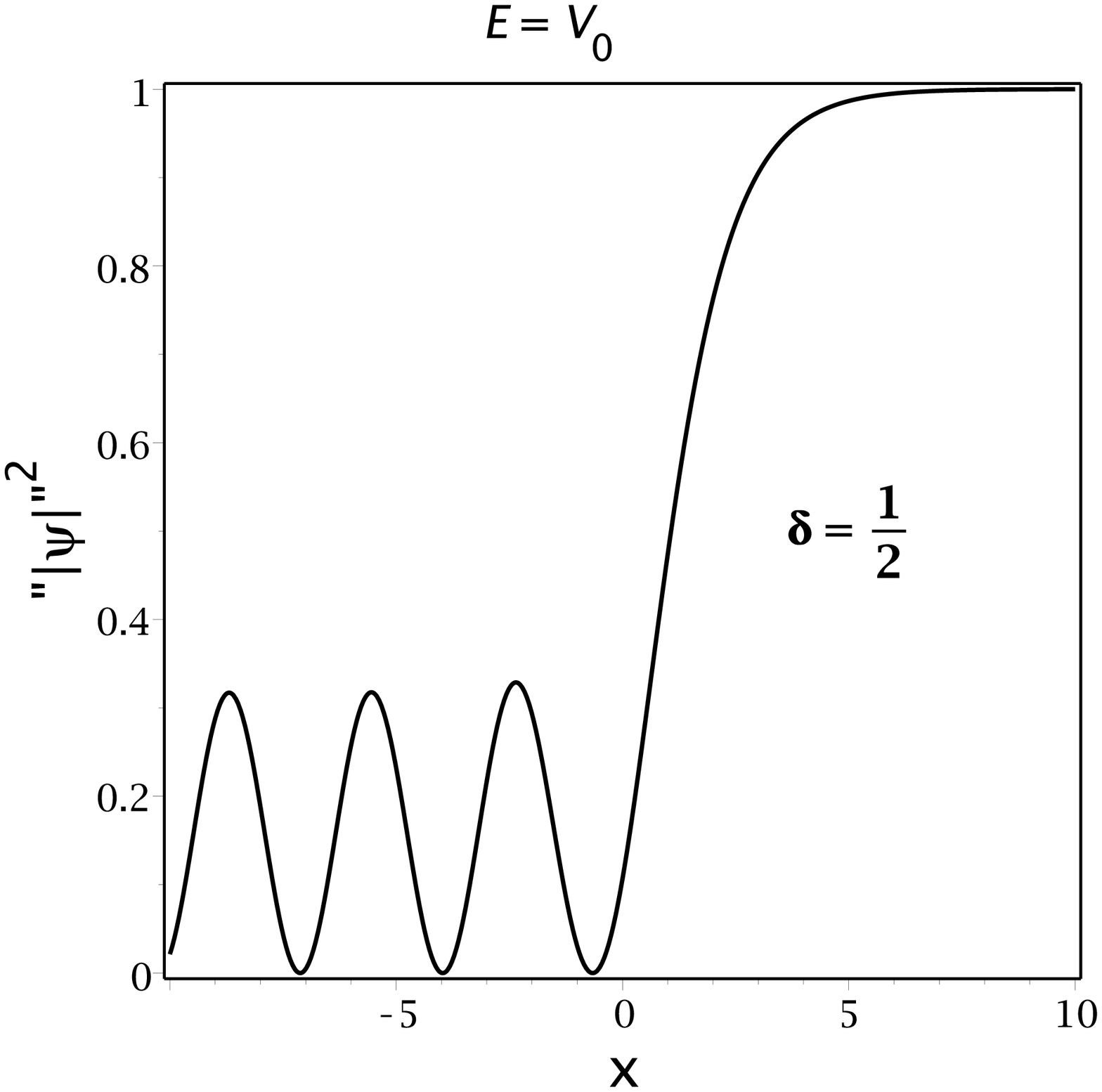}}\hspace{1cm}
{\includegraphics[width=6.8cm,height=6.8cm]{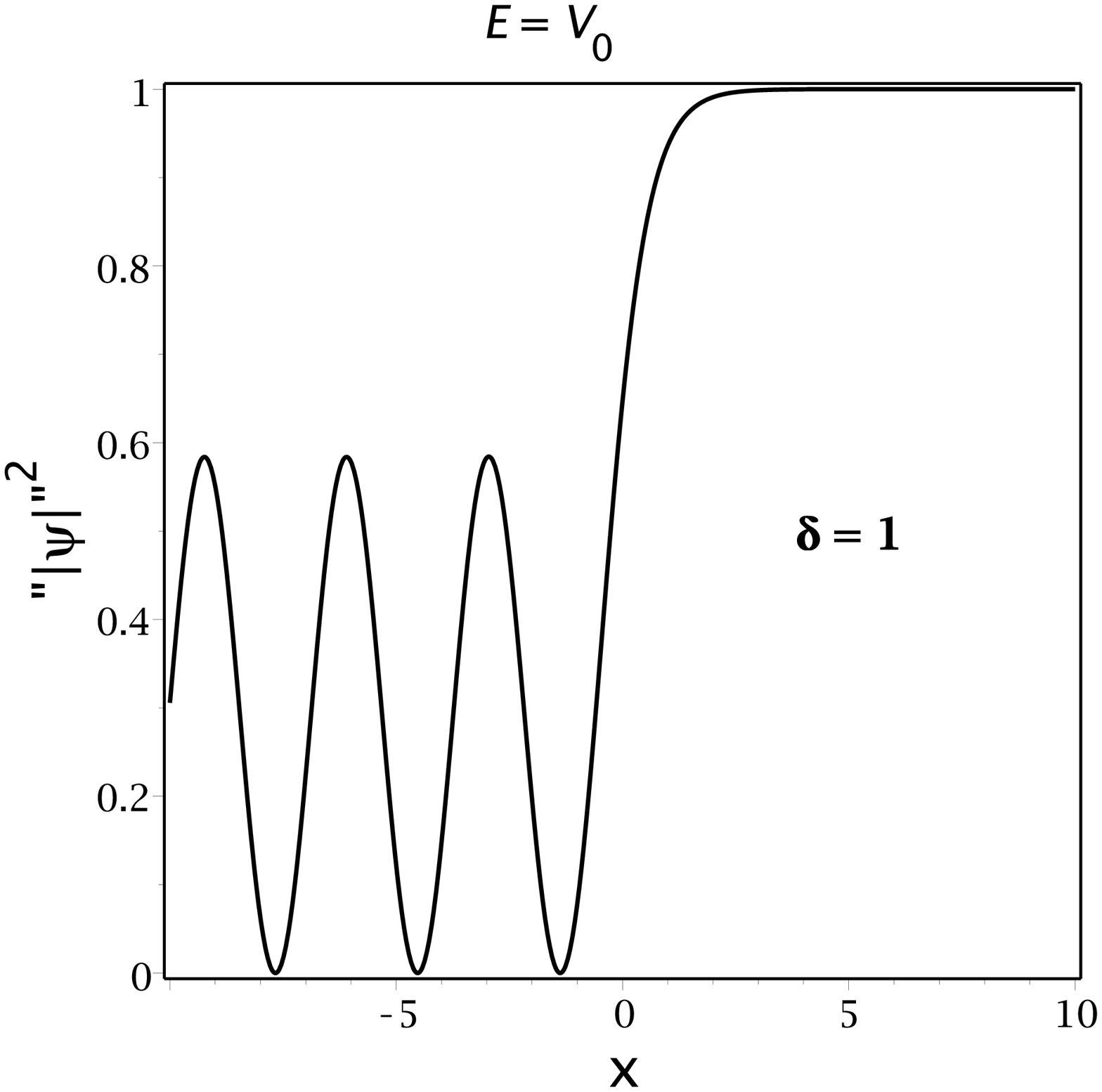}}
{\includegraphics[width=6.8cm,height=6.8cm]{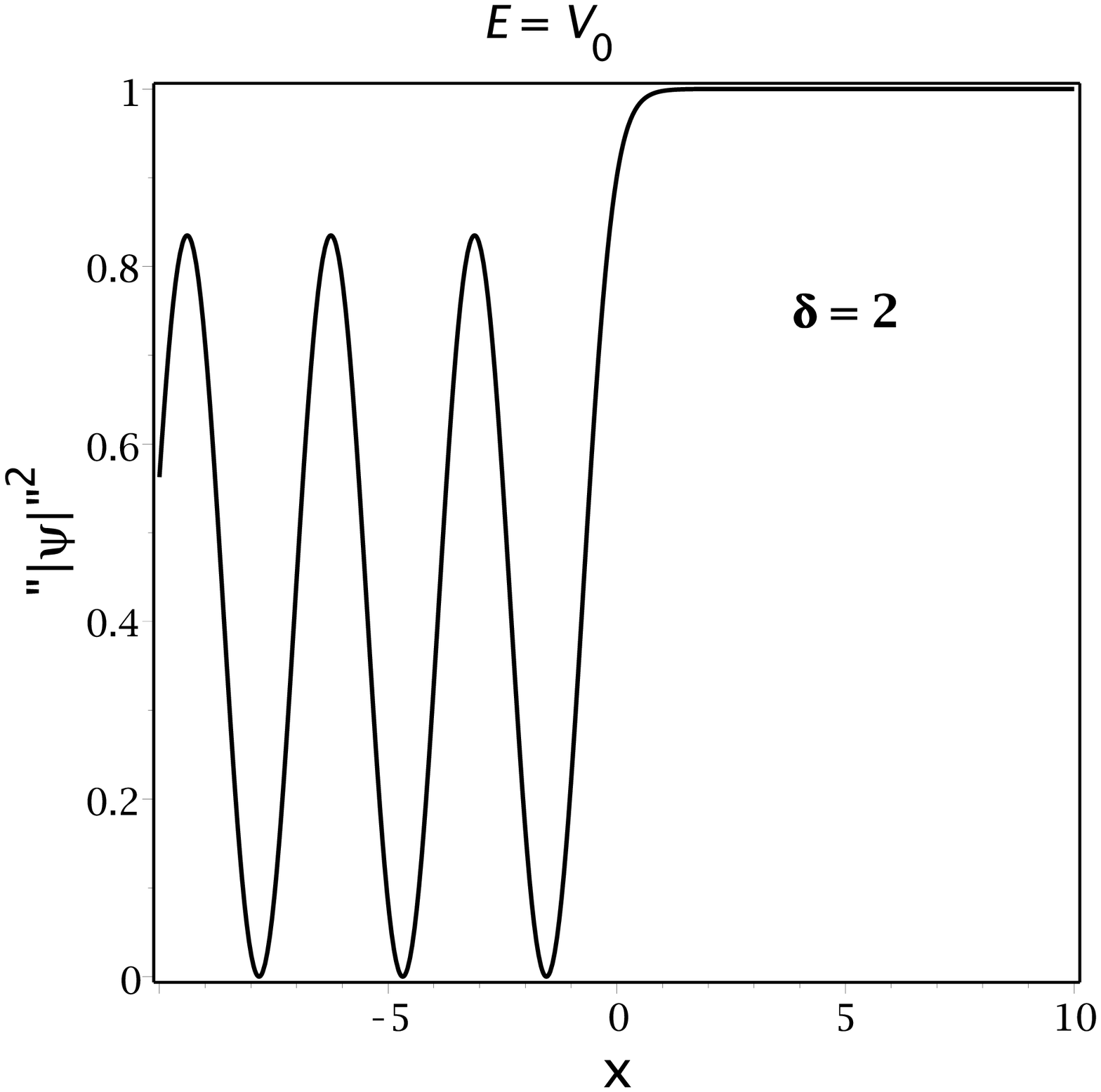}}\hspace{1cm}
{\includegraphics[width=6.8cm,height=6.8cm]{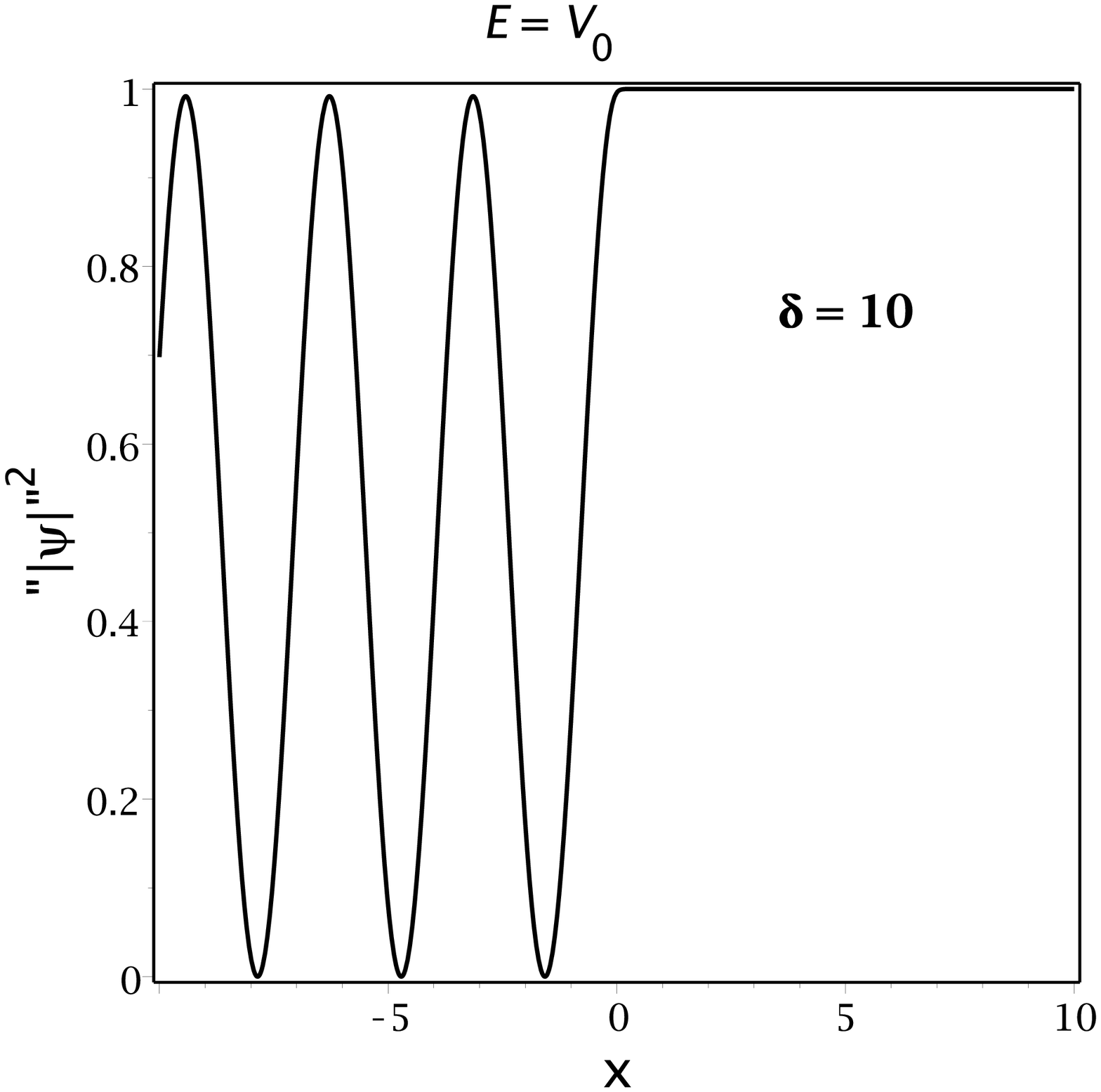}}
\caption{\label{fig_wf} M\'odulo quadrado da fun\c{c}\~ao de onda, $|\psi|^2$, para $E=V_0$ e para diversos valores de $\delta$, a saber, $\delta=1/2,~1,~2,$ e $10$.}
\end{figure}

\begin{figure}[h]
\center
{\includegraphics[width=6.8cm,height=6.8cm]{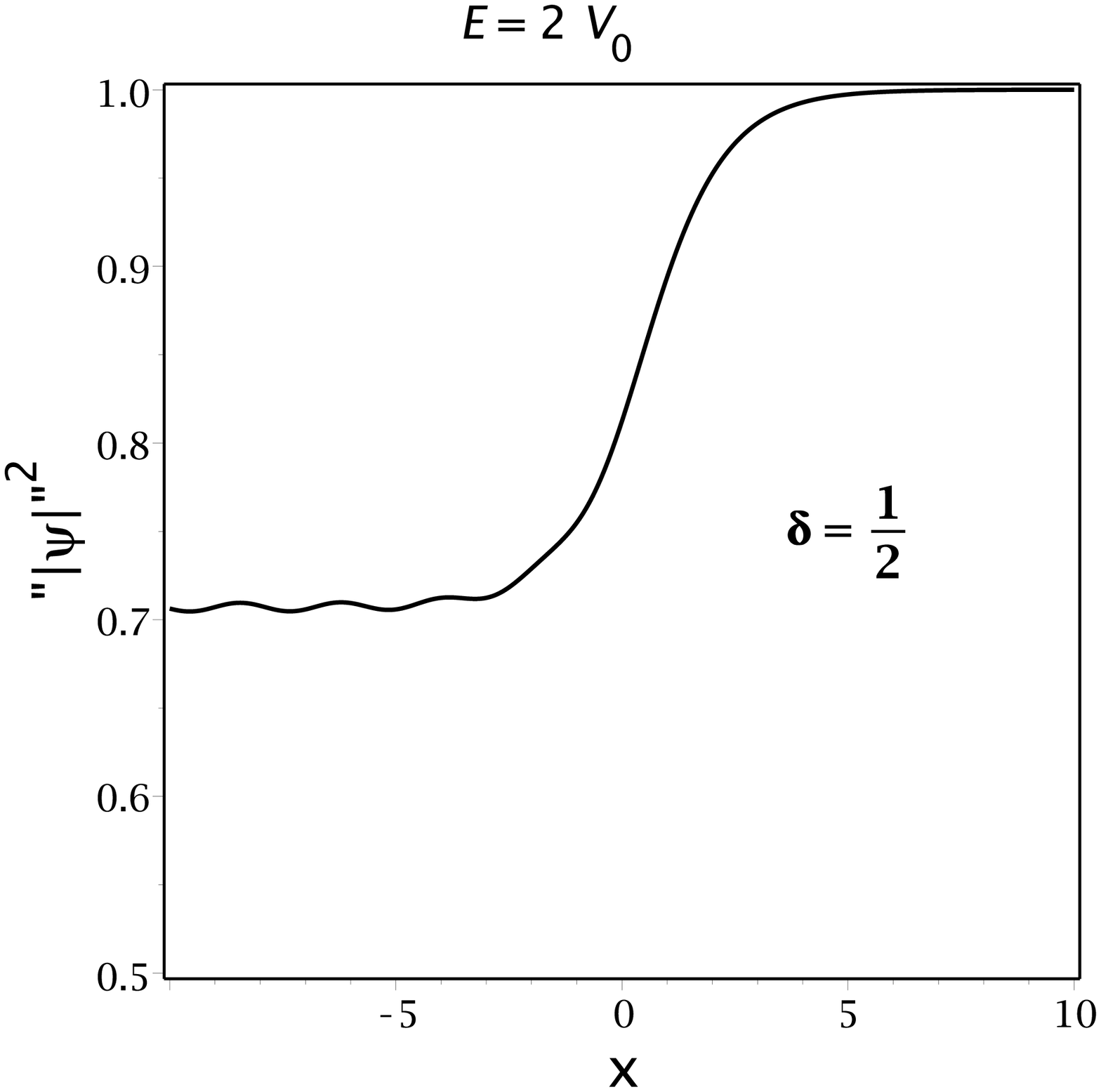}}\hspace{1cm}
{\includegraphics[width=6.8cm,height=6.8cm]{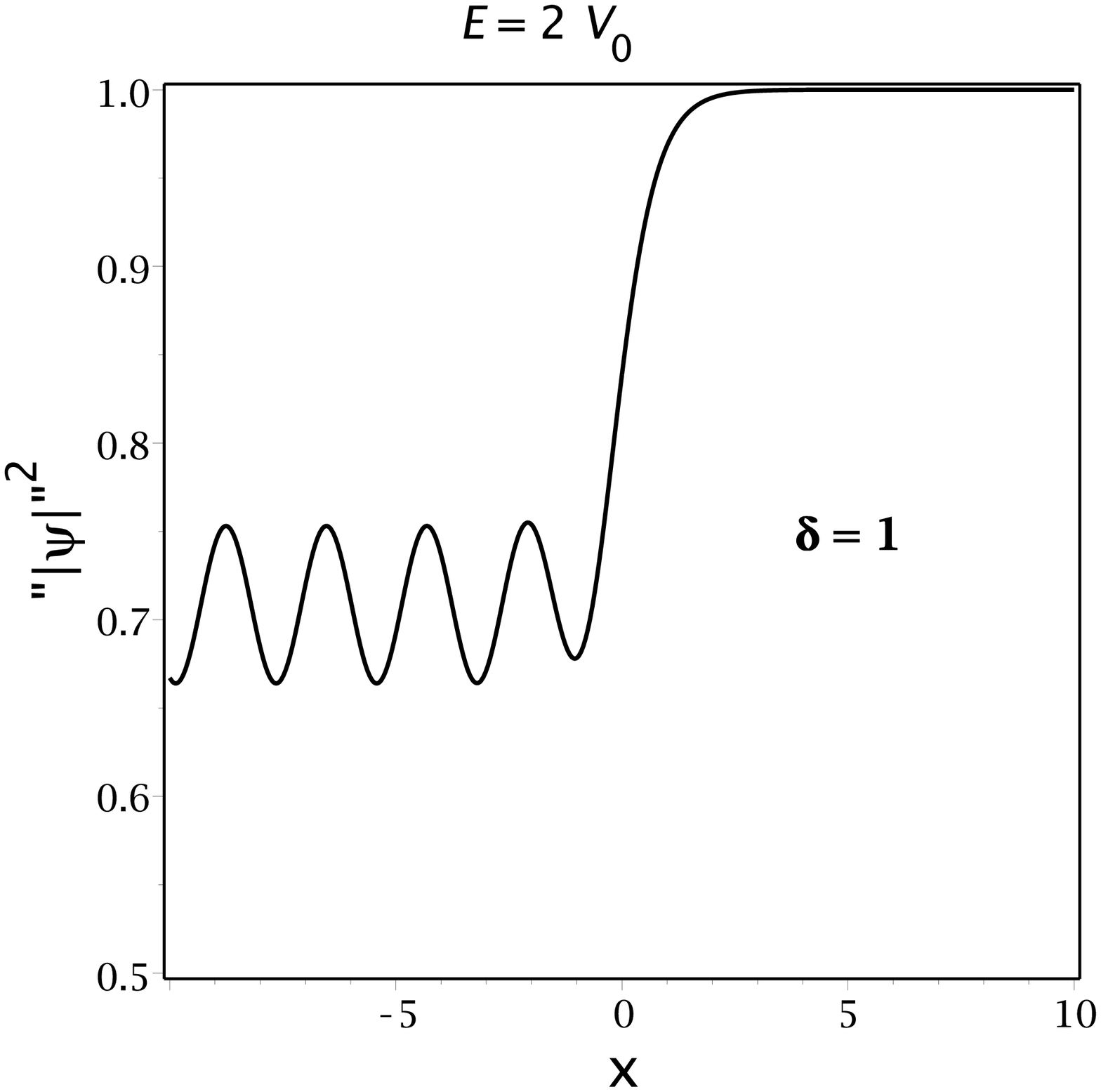}}
{\includegraphics[width=6.8cm,height=6.8cm]{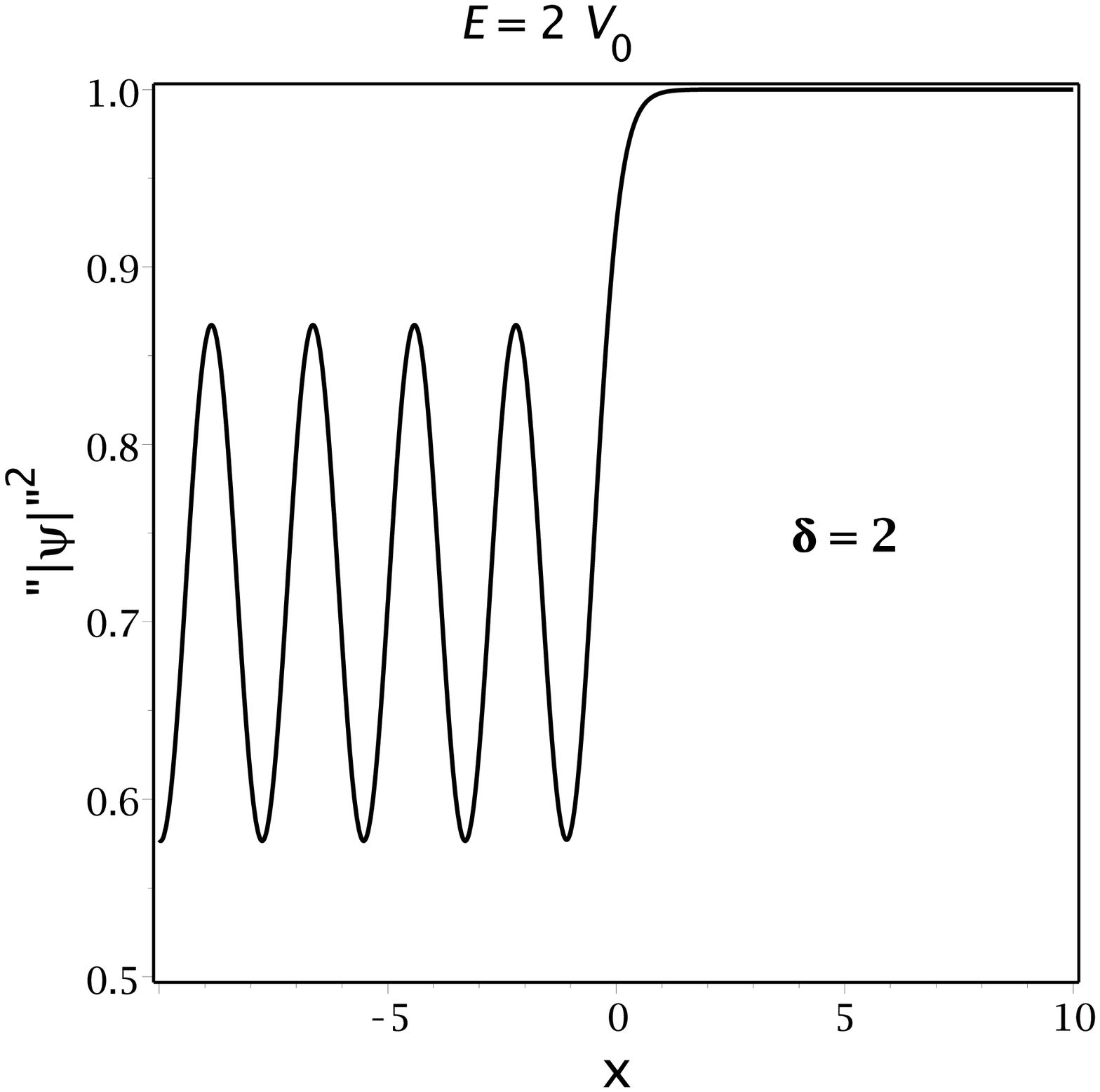}}\hspace{1cm}
{\includegraphics[width=6.8cm,height=6.8cm]{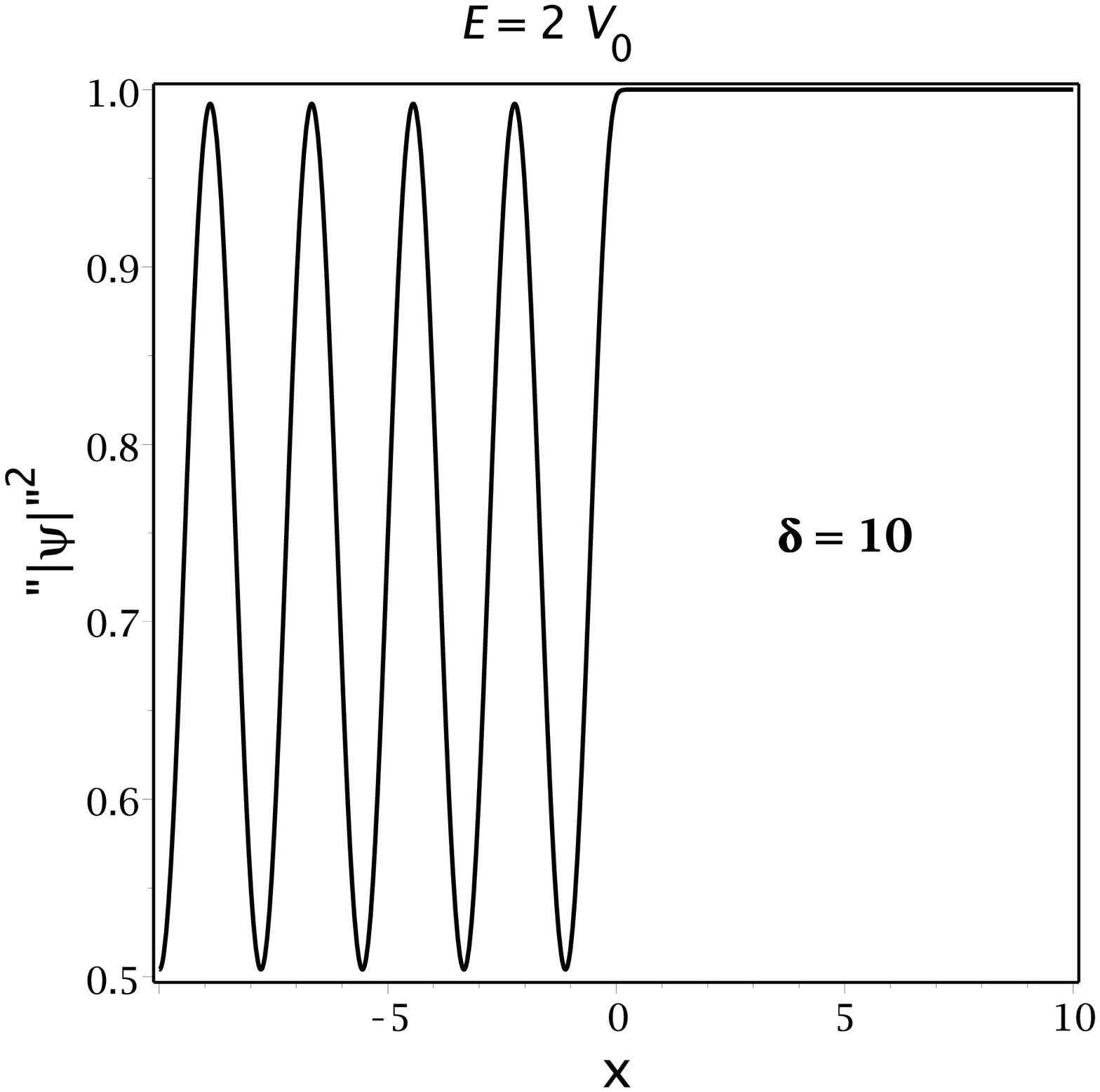}}
\caption{\label{fig_wf_E2V} M\'odulo quadrado da fun\c{c}\~ao de onda, $|\psi|^2$, para $E=2V_0$ e para diversos valores de $\delta$, a saber, $\delta=1/2,~1,~2,$ e $10$.}
\end{figure}

\noindent\hspace{1.5cm}A partir da solu\c{c}\~ao de onda acima, apesar de ela pr\'opia n\~ao satisfazer o limite assint\'otico imposto pela Eq. (\ref{assim_L}), podemos obter as ondas incidente e refletida. Para tanto, usamos uma rela\c{c}\~ao particularmente \'util entre fun\c{c}\~oes hipergeom\'etricas  \cite{ABRAMOWITZ}
\bea
{}_2F_1(a,b,c;z)=\frac{\Gamma(c)\Gamma(b-a)}{\Gamma(b)\Gamma(c-a)}(-z)^{-a}{}_2F_{1}(a, 1+a-c,1+a-b;z^{-1})\nonumber\\
+~\frac{\Gamma(c)\Gamma(a-b)}{\Gamma(a)\Gamma(c-b)}(-z)^{-b}{}_2F_{1}(b, 1+b-c,1+b-a;z^{-1}),
\eea
com $|arg(-z)|< \pi$, a fim de transformar a Eq. (\ref{eq_T}). Temos ent\~ao,
%
\bea
\psi(x) = A\, e^{i2\delta \mu x}(1+e^{2\delta x})\, {}_2F_1 \Big(1+i(\mu-\nu),1+i(\mu+\nu),1+2i\mu;-e^{2\delta x}\Big)~~~\nonumber\\
+ B\, e^{-i2\mu \delta x}(1+e^{2\delta x})\,{}_2F_1 \Big(1-i(\mu+\nu),1-i(\mu-\nu),1 -2i\mu;-e^{2\delta x}\Big).
\eea
%
onde 
\begin{subequations}
\bea 
A &\equiv& C_1 (-1)^{-i\nu} \frac{\Gamma(1-i2\nu)\Gamma(-i2\mu)}{\Gamma[1-i(\mu+\nu)]\Gamma[-i(\mu+\nu)]} \label{eq_A} \\
B &\equiv& C_1 (-1)^{-i\nu}\frac{\Gamma(1-i2\nu)\Gamma(i2\mu)}{\Gamma[1+i(\mu-\nu)]\Gamma[i(\mu-\nu)]}\label{eq_B}
\eea %
\end{subequations}
\noindent\hspace{1.5cm}Assim, neste caso, temos que as solu\c{c}\~oes fisicamente aceit\'aveis s\~ao aquelas equivalentes a ondas incidindo pela esquerda e que satisfazem as condi\c{c}\~oes  assint\'oticas adequadas, Eq. (\ref{assim_L}), e podem ser escritas explicitamente como uma combina\c{c}\~ao linear de uma onda incidente e outra refletida, ou seja,
\beq
\Psi_L(x)=A\,\psi_{inc}(x)+B\,\psi_{ref}(x), \label{psi_L}
\eeq
onde 
\bea
\psi_{inc}(x) \!\!&=&\!\!  e^{i2\delta \mu x}(1+e^{2\delta x})
{}_2F_1 \Big(1+i(\mu-\nu),1+i(\mu+\nu),1+2i\mu;-e^{2\delta x}\Big) \label{psi_inc}\\
\psi_{ref}(x)\!\!&=&\!\! e^{-i2\mu \delta x}(1+e^{2\delta x}){}_2F_1 \Big(1-i(\mu+\nu),1-i(\mu-\nu),1 -2i\mu;-e^{2\delta x}\Big),\label{psi_ref}
\eea
com $A$ e $B$ definidos acima. Abaixo, as Figs. (\ref{fig_wf}) e (\ref{fig_wf_E2V}) representam a densidade de probabilidade, $|\psi|^2$, para os casos onde $\mathcal{E}=\mathcal{V}_0$ e $\mathcal{E}=2\mathcal{V}_0$ e para alguns valores do par\^amtro de deforma\c{c}\~ao $\delta$.

\noindent\hspace{1.5cm} O leitor mais atento poderia se perguntar como uma \'unica solu\c{c}\~ao pode gerar duas outras solu\c{c}\~oes linearmente independentes (L. I.). Mas \'e esse exatamente o caso e pode ser conferido utilizando-se a transforma\c{c}\~ao $y=-e^{2\delta x}$ no lugar da Eq. (\ref{transy}). \'E interessante o leitor pesquisar um pouco sobre equa\c{c}\~oes hipergeom\'etricas e ver que, no caso da hipergeom\'etrica de Gauss, \'e poss\'ivel, por causa dos pontos singulares da equa\c{c}\~ao e, portanto, de suas simetrias, encontrar duas solu\c{c}\~oes L. I. para cada um dos tr\^es pontos singulares, ou seja, seis solu\c{c}\~oes. Ainda, por causa das simetrias da equa\c{c}\~ao descritas pelas transforma\c{c}\~oes de M\"obius, \'e poss\'ivel encontrar mais tr\^es solu\c{c}\~oes L. D. (linearmente dependentes) para cada solu\c{c}\~ao L. I. A equa\c{c}\~ao hipergeom\'etrica de Gauss possui, portanto, 24 solu\c{c}\~oes todas relacionadas pelas transforma\c{c}\~oes de M\"obius \cite{MAIER,ABRAMOWITZ,SLATER}
\\
\section{Coeficientes de transmiss\~ao e reflex\~ao}
\noindent\hspace{1.5cm}Como sugerem as solu\c{c}\~oes assint\'oticas, os coeficientes de refrex\~ao e transmiss\~ao podem ser obtidos diretamente das constantes $A$, $B$ e $C$ descritas acima. S\~ao dados, respectivamente, por $R=|B/A|^2$ e $T=(\nu/\mu)|C/A|^2$. Explicitamente, temos
\beq
R=\left|\frac{\Gamma(i2\mu)\,\Gamma[1-i(\mu+\nu)]\,\Gamma[-i(\mu+\nu)]} {\Gamma(-i2\mu)\,\Gamma[1+i(\mu-\nu)]\, \Gamma[i(\mu-\nu)]}\right|^2 \label{reflex}
\eeq
\beq
T= \frac{\nu}{\mu}\left|\frac{\Gamma[1-i(\mu+\nu)]\,\Gamma[-i(\mu+\nu)]}{\Gamma(1-i2\nu)\,\Gamma(-i2\mu)}\right|^2 \label{trans}
\eeq
\noindent\hspace{1.5cm}Podemos simplificar as express\~oes acima utilizando, primeiro, o fato de que o m\'odulo quadrado de grandezas conjugadas complexas uma da outra s\~ao iguais. Assim, $|\Gamma(i2\mu)|^2=|\Gamma(-i2\mu)|^2$. Segundo, podemos usar as express\~oes \cite{FLUGGE, ABRAMOWITZ}
\begin{subequations}
\bea
\Gamma(z)=\frac{1}{z} \Gamma(1+z)\label{gama1}\\
|\Gamma(1+i\eta)|^2=\frac{\pi\eta}{\sinh(\pi\eta)},~\label{gama2}
\eea
onde $\eta \in \mathbb{R}$.
\end{subequations}
As express\~oes dos coeficientes de reflex\~ao e transmiss\~ao s\~ao reescritas, usando a primeira das express\~oes acima, como
\bea
R&\!\!\!=\!\!\!&\frac{(\mu-\nu)^2}{(\mu+\nu)^2}\left|\frac{\Gamma[1-i(\mu+\nu)]}{\Gamma[1+i(\mu+\nu)]}\right|^4\\
T&\!\!\!=\!\!\!&\frac{4\mu\nu}{(\mu+\nu)^2}\frac{\left|\Gamma[1-i(\mu+\nu)]\right|^4}{\left|\Gamma(1-i2\mu) \Gamma(1-i2\nu)\right|^2}
\eea
Utilizando a segunda express\~ao, Eq. (\ref{gama2}), podemos reescrever os coeficientes em termos de fun\c{c}\~oes hiperb\'olicas como 
\bea
R &\!\!\!=\!\!\!&\frac{\sinh^2\!\pi(\mu-\nu)}{\sinh^2\!\pi(\mu+\nu)}\\
T&\!\!\!=\!\!\!&\frac{\sinh(2\pi\mu)\sinh(2\pi\nu)}{\sinh^2\!\pi(\mu+\nu)}
\eea 
A partir destas express\~oes, n\~ao \'e dif\'icil verificar a rela\c{c}\~ao $R+T=1$.

\noindent\hspace{1.5cm}A partir desse ponto, podemos averiguar com mais facilidade o caso limite quando $\delta \gg 1$. Este \'e o caso quando o potencial, dado pela Eq. (\ref{pot}), tende para o potencial degrau e os param\^etros $\mu$ e $\nu$, dados pela Eq. (\ref{munu}), se tornam muito pequenos. Podemos pois considerar $\sinh(x)\approx x$, o que produz as seguintes rela\c{c}\~oes para os coeficientes de reflex\~ao e transmiss\~ao, a saber,
\bea
R \approx \frac{(k-\ell)^2}{(k+\ell)^2}\\
T \approx \frac{4k\ell}{(k+\ell)^2}
\eea
com $k^2=\mathcal{E}$ e $\ell^2=\mathcal{E}-\mathcal{V}_0$ (exatamente como nos casos assint\'oticos $x\rightarrow \pm \infty$).%
\section{Penetra\c{c}\~ao na barreira de potencial: caso $E<V_0$}
\noindent\hspace{1.5cm}Para casos onde a energia \'e menor que $V_0$, as solu\c{c}\~oes \`a direita deixam  de ser do tipo part\'icula livre e passam a se comportar como a Eq. (\ref{pb}). Esta \'e a raz\~ao f\'isica de considerarmos, desta vez, a segunda solu\c{c}\~ao da Eq. (\ref{eq_psix}). Temos ent\~ao,
\bea
\Psi_{p} (x) &=& \tilde{D} \psi_\kappa(x) \label{eq_p}
\eea
onde
\bea
\psi_{\kappa} (x) &=& \,e^{-\kappa x}(1+e^{-2\delta x})\,{}_2F_1 \Big(1+\frac{\kappa}{2\delta}+i\mu,1+\frac{\kappa}{2\delta}-i\mu,1+\frac{\kappa}{\delta};-e^{-2\delta x}\Big),\label{eq_k}
\eea
com $\kappa=\sqrt{\mathcal{V}_0-\mathcal{E}}$ e $\tilde{D}=C_2 (-1)^{\kappa/2\delta}$. As solu\c{c}\~oes \`a esquerda s\~ao semelhantes \`as Eqs. (\ref{psi_inc}) e (\ref{psi_ref}), exceto que agora devemos substituir $i\nu=\kappa/2\delta$ nas respectivas express\~oes, ou seja,
\begin{subequations}
\bea 
\tilde{A} &\equiv& C_1 (-1)^{-\kappa/2\delta} \frac{\Gamma(1-\kappa/\delta)\Gamma(-i2\mu)}{\Gamma[1-\kappa/2\delta-i\mu]\Gamma[-\kappa/2\delta-i\mu]} \label{eq_A_2} \\
\tilde{B} &\equiv& C_1 (-1)^{-\kappa/2\delta}\frac{\Gamma(1-\kappa/\delta)\Gamma(i2\mu)}{\Gamma[1-\kappa/2\delta+i\mu]\Gamma[-\kappa/2\delta +i\mu)]}\label{eq_B_2}
\eea %
\end{subequations}
e a solu\c{c}\~ao \`a esquerda da origem fica
\beq
\tilde{\Psi}_{L}(x)=\tilde{A}\,\tilde{\psi}_{inc}(x)+\tilde{B}\,\tilde{\psi}_{ref}(x), \label{psi_L_2}
\eeq
sendo que as ondas incidente e refletida s\~ao agora dadas por
\begin{subequations}
\bea
\tilde{\psi}_{inc}(x) \!\!&=&\!\!  e^{i2\delta \mu x}(1+e^{2\delta x})
{}_2F_1 \Big(1-\kappa/2\delta+i\mu,1+\kappa/2\delta + i\mu,1+2i\mu;-e^{2\delta x}\Big) \label{psi_inc_2}\\
\tilde{\psi}_{ref}(x)\!\!&=&\!\! e^{-i2\mu \delta x}(1+e^{2\delta x}){}_2F_1 \Big(1-\kappa/2\delta-i\mu,1+\kappa/2\delta-i\mu,1 -2i\mu;-e^{2\delta x}\Big).\label{psi_ref_2}
\eea
\end{subequations}
O coeficiente de reflex\~ao pode ser obtido diretamente da Eq. (\ref{reflex}), trocando $i\nu\rightarrow \kappa/2\delta$, ou utilizando-se as Eqs. (\ref{eq_A_2}) e (\ref{eq_B_2}).  Temos ent\~ao,
\beq \label{reflexao_total}
R=\left|\frac{\tilde{B}}{\tilde{A}} \right|^2= \left|\frac{\Gamma(i2\mu)\,\Gamma(1 -\kappa/2\delta-i\mu)\,\Gamma(-\kappa/2\delta-i\mu)}{\Gamma(-i2\mu)\,\Gamma(1-\kappa/2\delta+i\mu)\,\Gamma(-\kappa/2\delta+i\mu)}\right|^2\!\!\!.
\eeq
\begin{figure}[h]
\center
{\includegraphics[width=7cm,height=7cm]{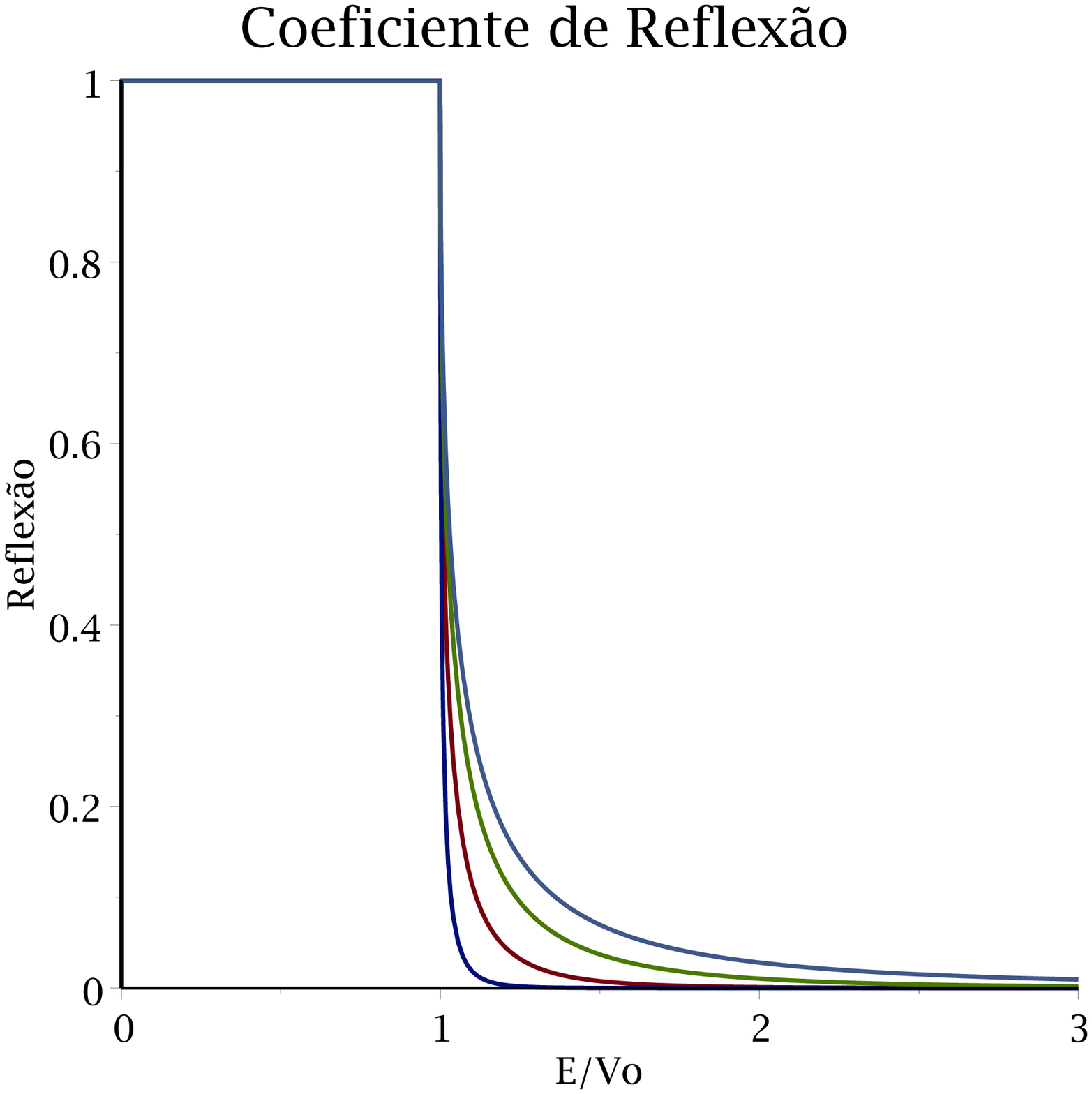}}\hspace{1cm}
{\includegraphics[width=7cm,height=7cm]{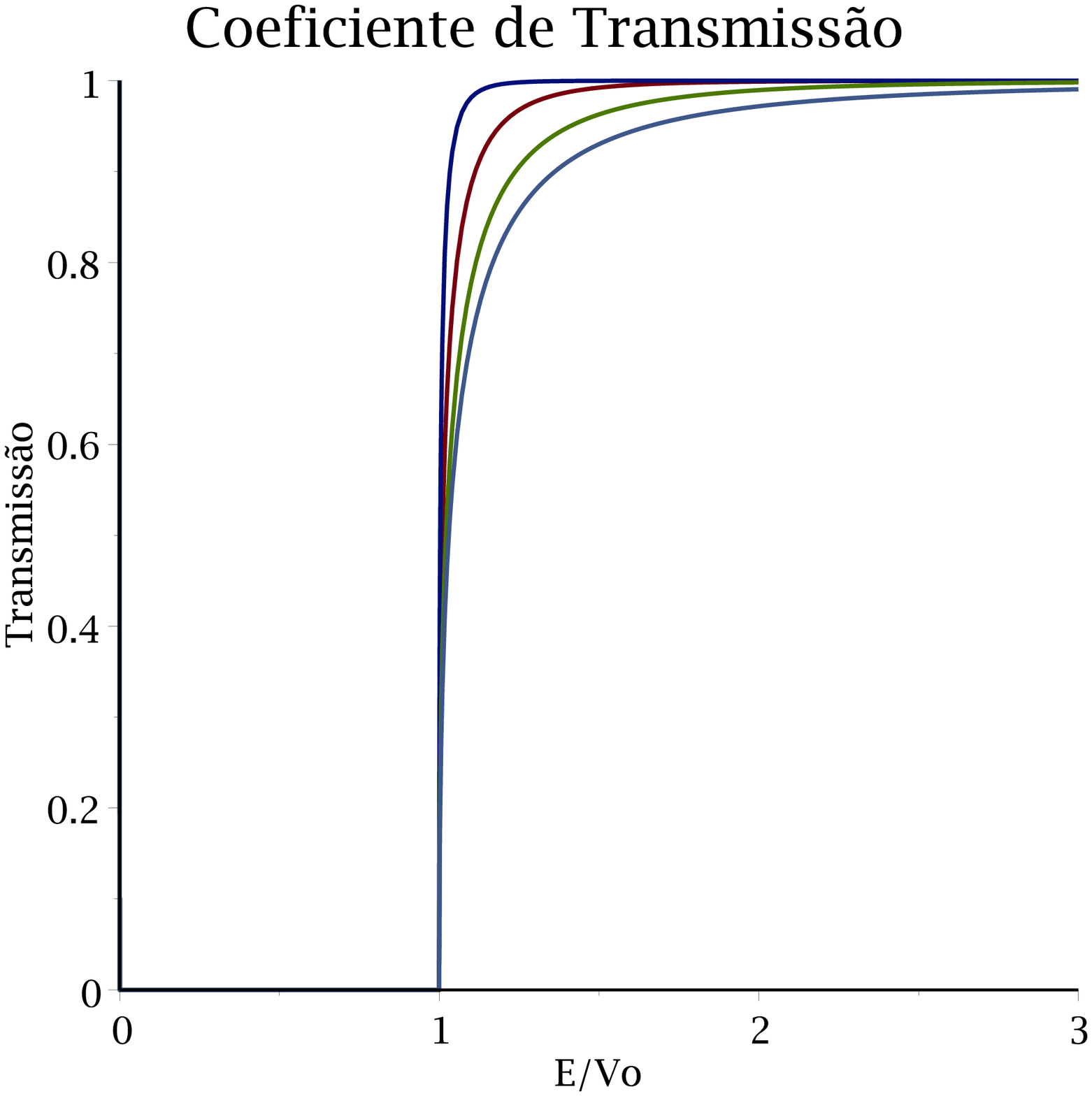}}
\caption{\label{fig_coef} Gr\'aficos dos coeficientes de reflex\~ao e transmiss\~ao em fun\c{c}\~ao da raz\~ao $\mathcal{E}/\mathcal{V}_0$, onde $\mathcal{E}=(2m/\hbar^2)E$ e $\mathcal{V}_0=(2m/\hbar^2)V_0$, para diferentes valores de $\delta$. Em cada gr\'afico, as curvas da esquerda para a direita correspondem a, respectivamente, $\delta=1/2,~1,~2,~10$.}
\end{figure}

\noindent\hspace{1.5cm}Podemos facilmente constatar que o denominador da equa\c{c}\~ao acima \'e o complexo conjugado do numerador. Como o m\'odulo quadrado de um n\'umero complexo \'e igual ao m\'odulo quadrado de seu conjugado, isso implica que o coeficiente de reflex\~ao \'e identicamente igual a 1. Esse resultado mostra que, apesar da fun\c{c}\~ao de onda penetrar na barreira, inexoravelmente toda a onda ser\'a refletida de volta. Esse resultado imp\~oe, por conseguinte, que o coeficiente de transmiss\~ao seja nulo.

\noindent\hspace{1.5cm}Na Fig. (\ref{fig_coef}), mostramos os coeficientes de reflex\~ao e transmiss\~ao em fun\c{c}\~ao da raz\~ao da energia relativa \`a altura m\'axima da barreira de potencial, $\mathcal{E}/\mathcal{V}_0$, para diversos valores do par\^ametro $\delta$ respons\'avel por deformar a barreira de potencial. 

\noindent\hspace{1.5cm}O pr\'oximo passo \'e analisar as fun\c{c}\~oes de onda para os valores $\mathcal{E}<\mathcal{V}_0$. Os estados de energia nesse caso tamb\'em s\~ao cont\'inuos e as solu\c{c}\~oes est\~ao representadas acima, nas Eqs. (\ref{eq_p}) e  (\ref{psi_L_2}). Uma vez que a constante $\tilde{D}$ na Eq. (\ref{eq_p}) \'e desconhecida, podemos usar as condi\c{c}\~oes de continuidade das fun\c{c}\~oes e de suas derivadas na origem para determin\'a-la. De fato, \'e f\'acil mostrar que 
\beq
\tilde{D}=\frac{\tilde{\psi}_{inc}(0) \tilde{\psi}'_{ref}(0)-\tilde{\psi}'_{inc}(0)\tilde{\psi}_{ref}(0)}{\tilde{\psi}_\kappa(0)\tilde{\psi}'_{ref}(0)-\tilde{\psi}'_\kappa(0) \tilde{\psi}_{ref}(0)} \tilde{A}\label{dtil}
\eeq
Da mesma forma, podemos usar essas mesmas condi\c{c}\~oes de continuidade para escrever $\tilde{B}$ em fun\c{c}\~ao de $\tilde{A}$. A rela\c{c}\~ao \'e dada abaixo.
\beq
\tilde{B}=\frac{\tilde{\psi}_\kappa(0) \tilde{\psi}_{inc}'(0) -\tilde{\psi}'_\kappa(0) \tilde{\psi}_{inc}(0)}{\tilde{\psi}'_{\kappa}(0)\tilde{\psi}_{ref}(0)-\tilde{\psi}_\kappa(0)\tilde{\psi}'_{ref}(0)} \tilde{A}\label{btil}
\eeq

\noindent\hspace{1.5cm}\'E interessante analisar o comportamento das solu\c{c}\~oes para $\delta \gg 1$, como fizemos na se\c{c}\~ao anterior. Podemos analisar a pr\'opria equa\c{c}\~ao de Schr\"odinger, que pode fornecer as solu\c{c}\~oes bem conhecidas do potencial degrau, ou podemos averiguar como se comportam as solu\c{c}\~oes nesse limite. Essencialmente, a an\'alise do comportamento das solu\c{c}\~oes se resume a analisar a fun\c{c}\~ao hipergeom\'etrica de Gauss na origem. Para $c\neq 0,-1,-2,\dots$, $F(a,b,c;0)=1$. 

\noindent\hspace{1.5cm}Portanto, para $x$ positivo e $\delta \gg 1$, a Eq. (\ref{eq_k}) pode ser aproximada por  $\tilde{\psi}_k(x) \approx e^{-\kappa x}$. Da mesma forma, para as Eqs. (\ref{psi_inc_2}) e (\ref{psi_ref_2}), quando $x$ \'e nagativo e $\delta \gg 1$, temos $\tilde{\psi}_{inc}(x)\approx e^{ikx}$ e $\tilde{\psi}_{ref}(x)\approx e^{-ikx}$, respectivamente. Desta forma, as Eqs. (\ref{dtil}) e (\ref{btil}) poder ser aproximadas por
\bea
\tilde{D}\approx \frac{2k}{k+i\kappa}\tilde{A}\\
\tilde{B}\approx \frac{k-i\kappa}{k+i\kappa}\tilde{A},
\eea
que equivale, novamente, \`as rela\c{c}\~oes das constantes para o potencial degrau, como esperado.
\begin{figure}[ht]
\center
{\includegraphics[width=6.8cm,height=6.8cm]{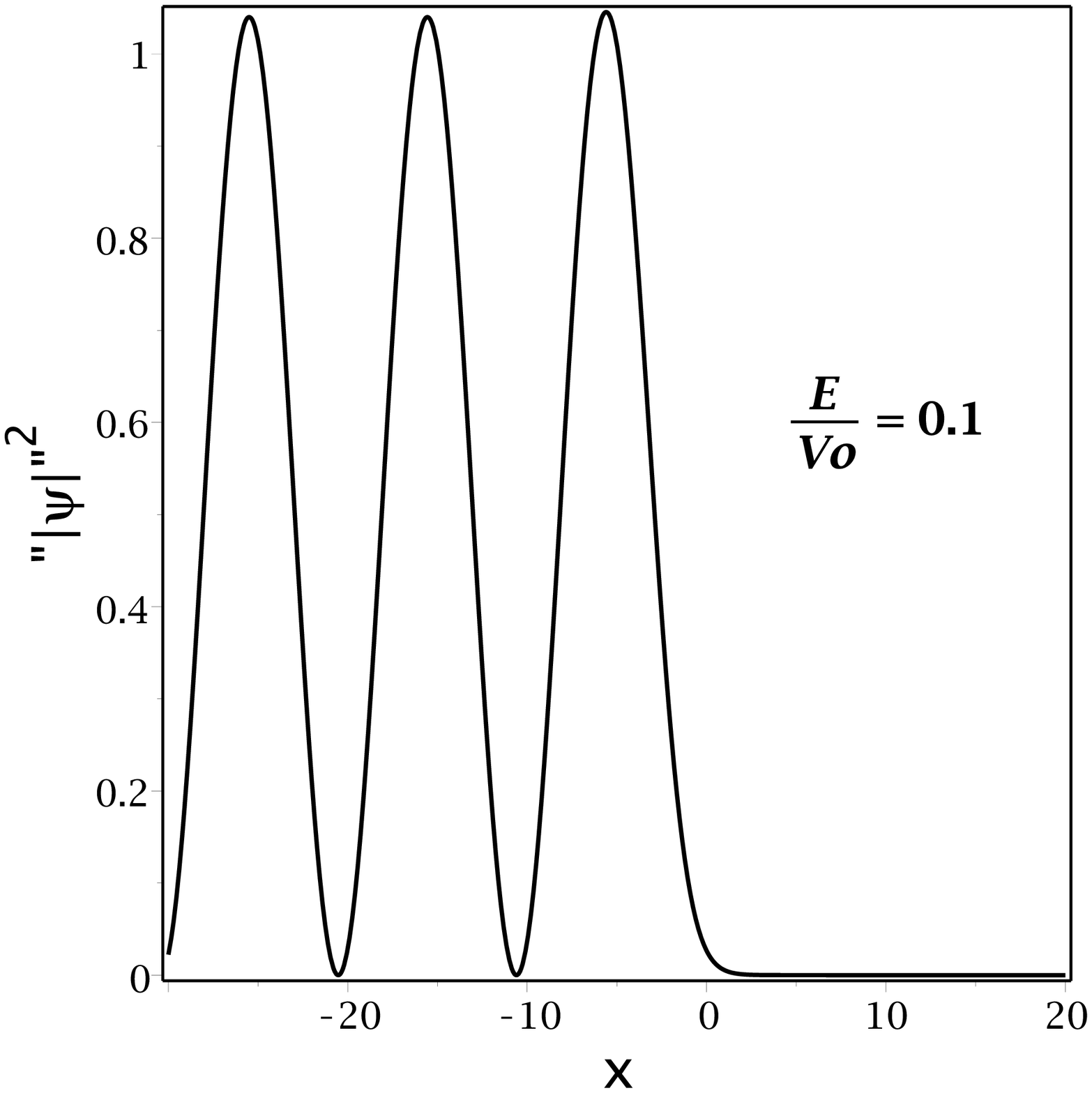}}\hspace{1cm}
{\includegraphics[width=6.8cm,height=6.8cm]{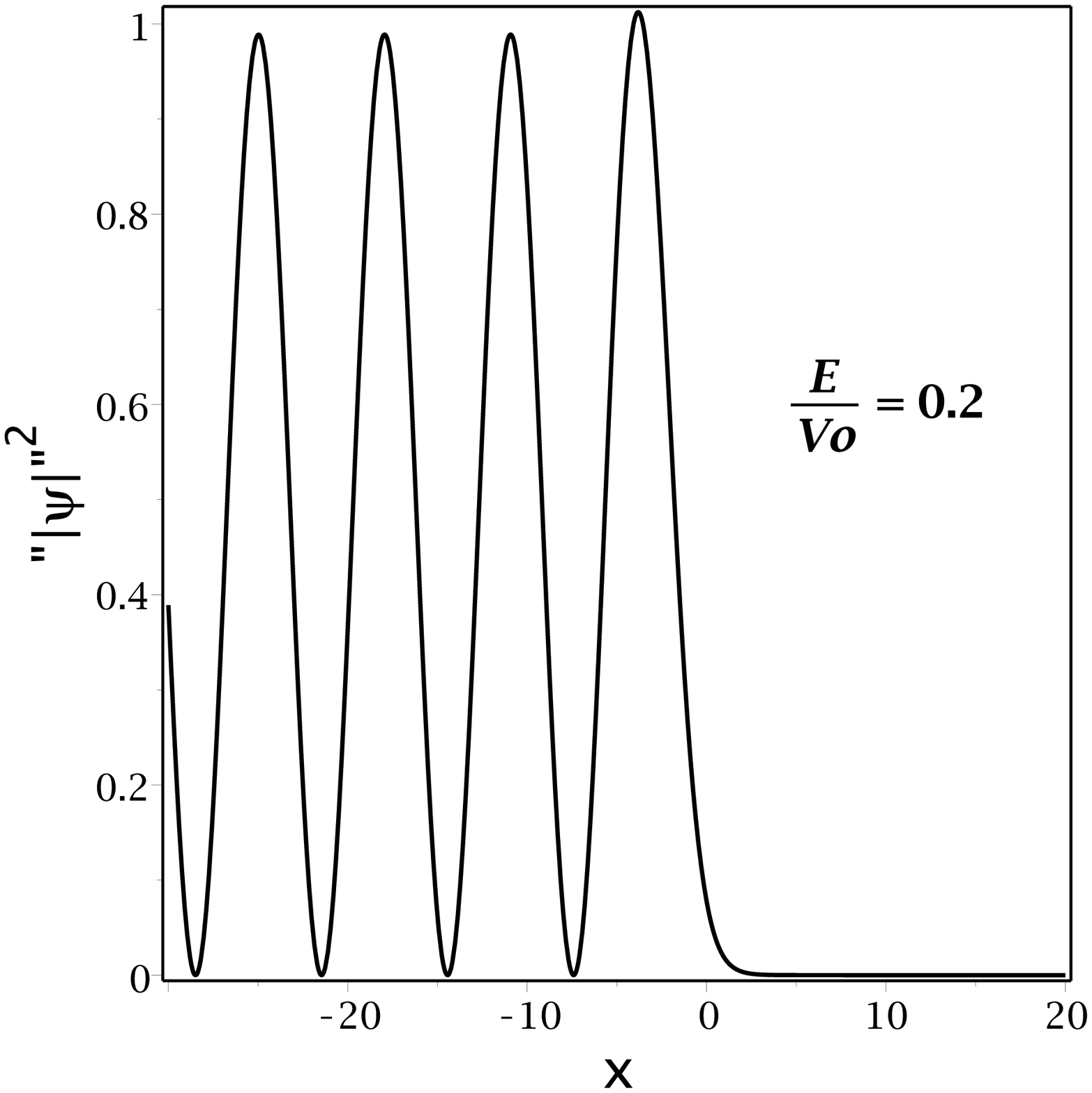}}
{\includegraphics[width=6.8cm,height=6.8cm]{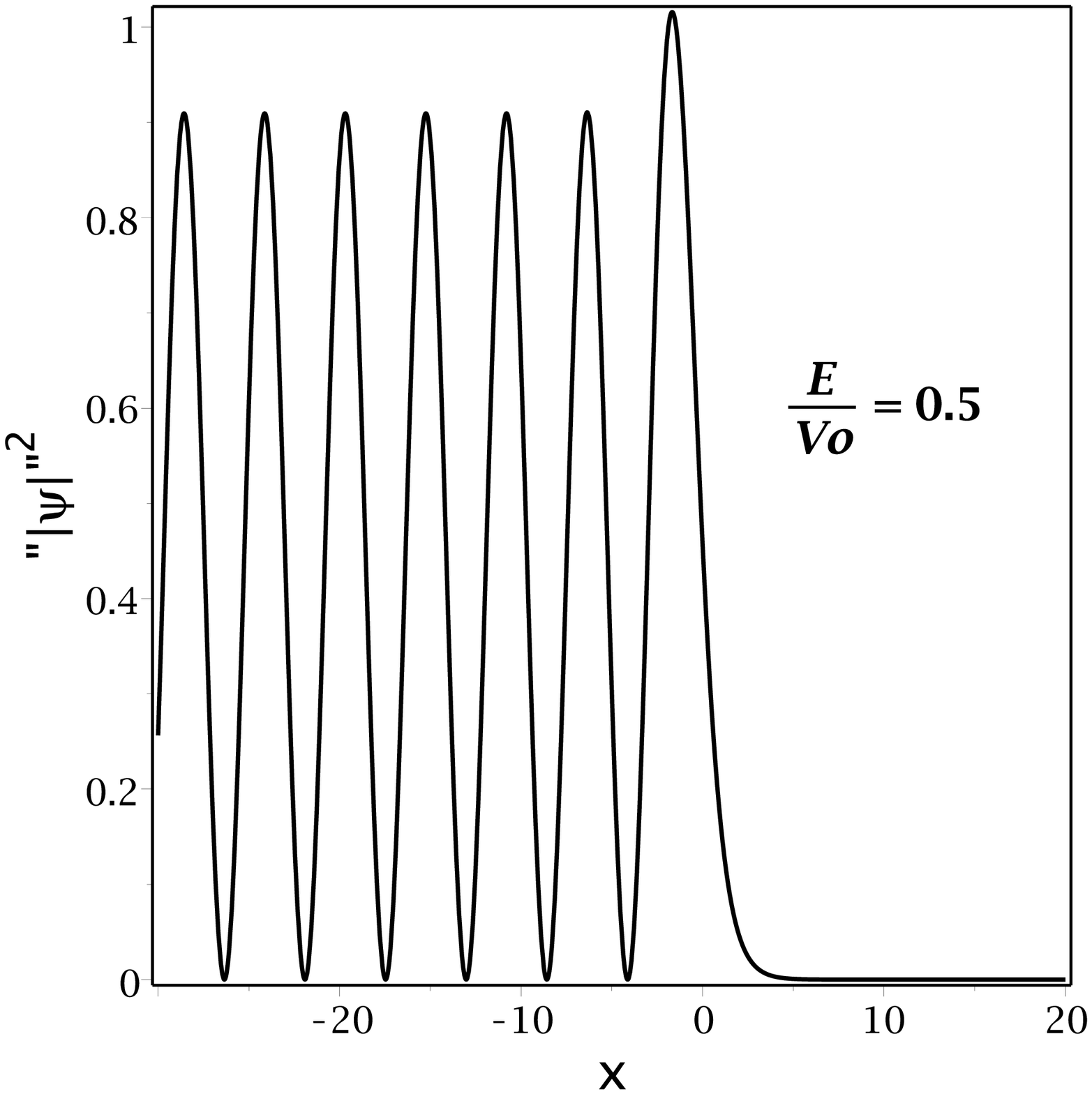}}\hspace{1cm}
{\includegraphics[width=6.8cm,height=6.8cm]{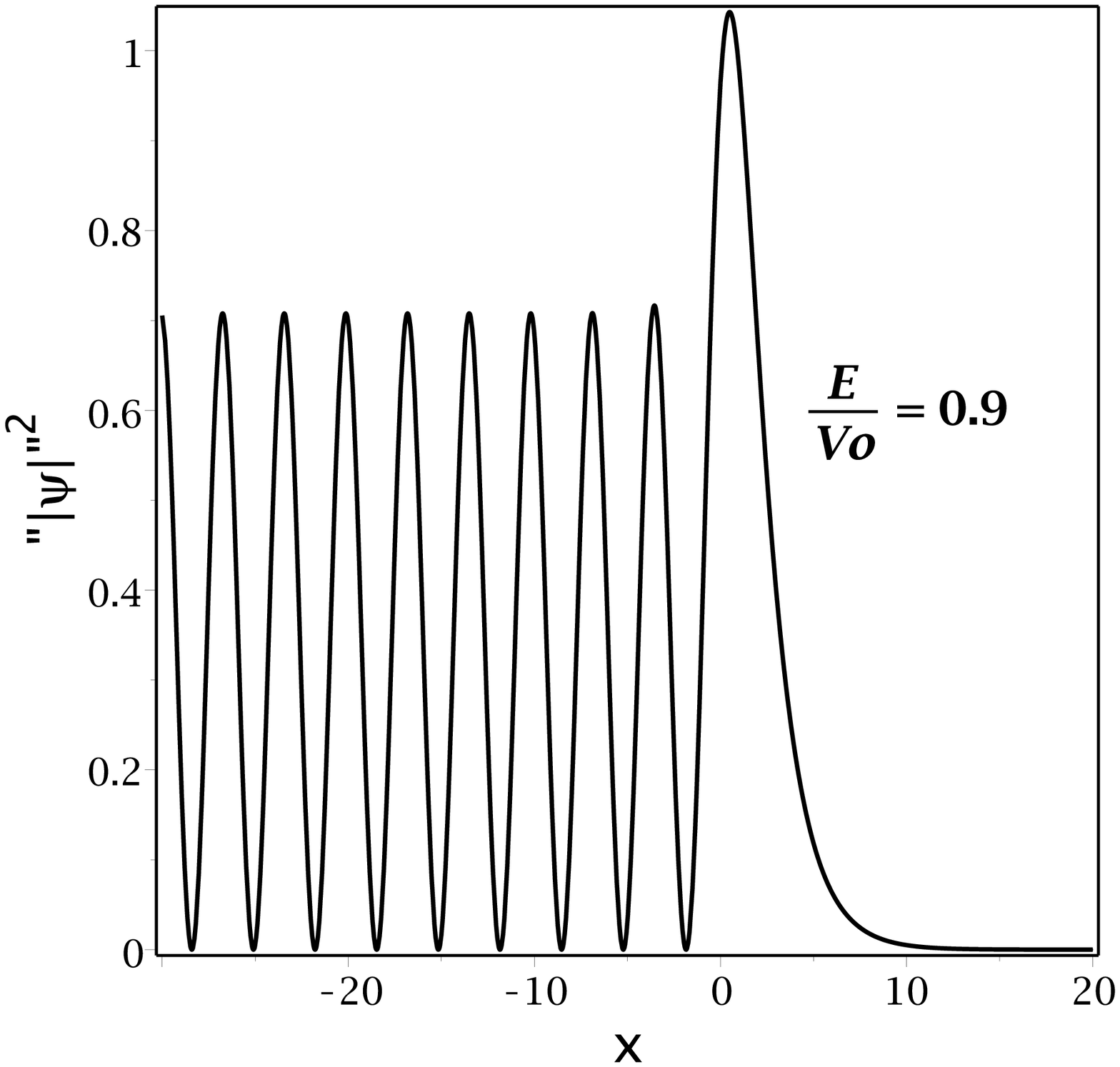}}
\caption{\label{fig_wf_E_menor_V}M\'odulo quadrado da fun\c{c}\~ao de onda, $|\psi|^2$, para energia abaixo da altura m\'axima da barreira de potencial, $\mathcal{E}/\mathcal{V}_0=0.1,~0.2,~0.5,~0.9$, para $\delta=1/2$. }
\end{figure}\\

\noindent\hspace{1.5cm}Escolhendo convenientemente $C_1(-1)^{-k/2\delta}$ na Eq. (\ref{eq_A_2}), apresentamos as figuras com conjuntos de gr\'aficos para alguns valores de $\mathcal{E}$ e de $\delta$. O primeiro conjunto, Fig. (\ref{fig_wf_E_menor_V}), representa a densidade de probabilidade para um perfil mais suave do potencial ($\delta=1/2$). Note-se o aumento relativo da densidade de probabilidade quando o valor da energia se aproxima da altura m\'axima do potencial, sendo inclusive mais prov\'avel encontrar a part\'icula em um intervalo um pouco mais \`a direita da origem, o que n\~ao acontece para o potencial degrau. 

\begin{figure}[ht]
\center
{\includegraphics[width=6.8cm,height=6.8cm]{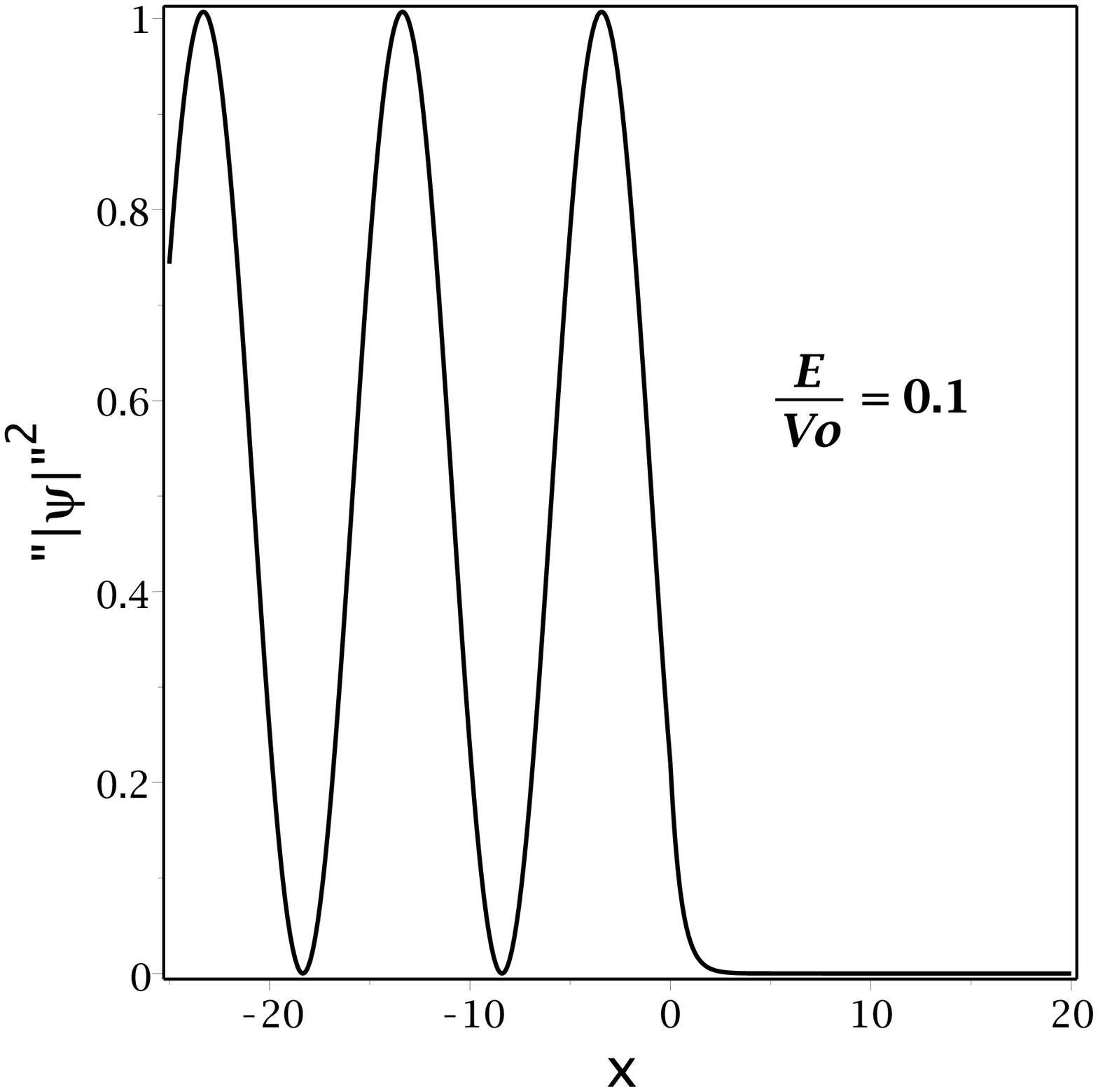}}\hspace{1cm}
{\includegraphics[width=6.8cm,height=6.8cm]{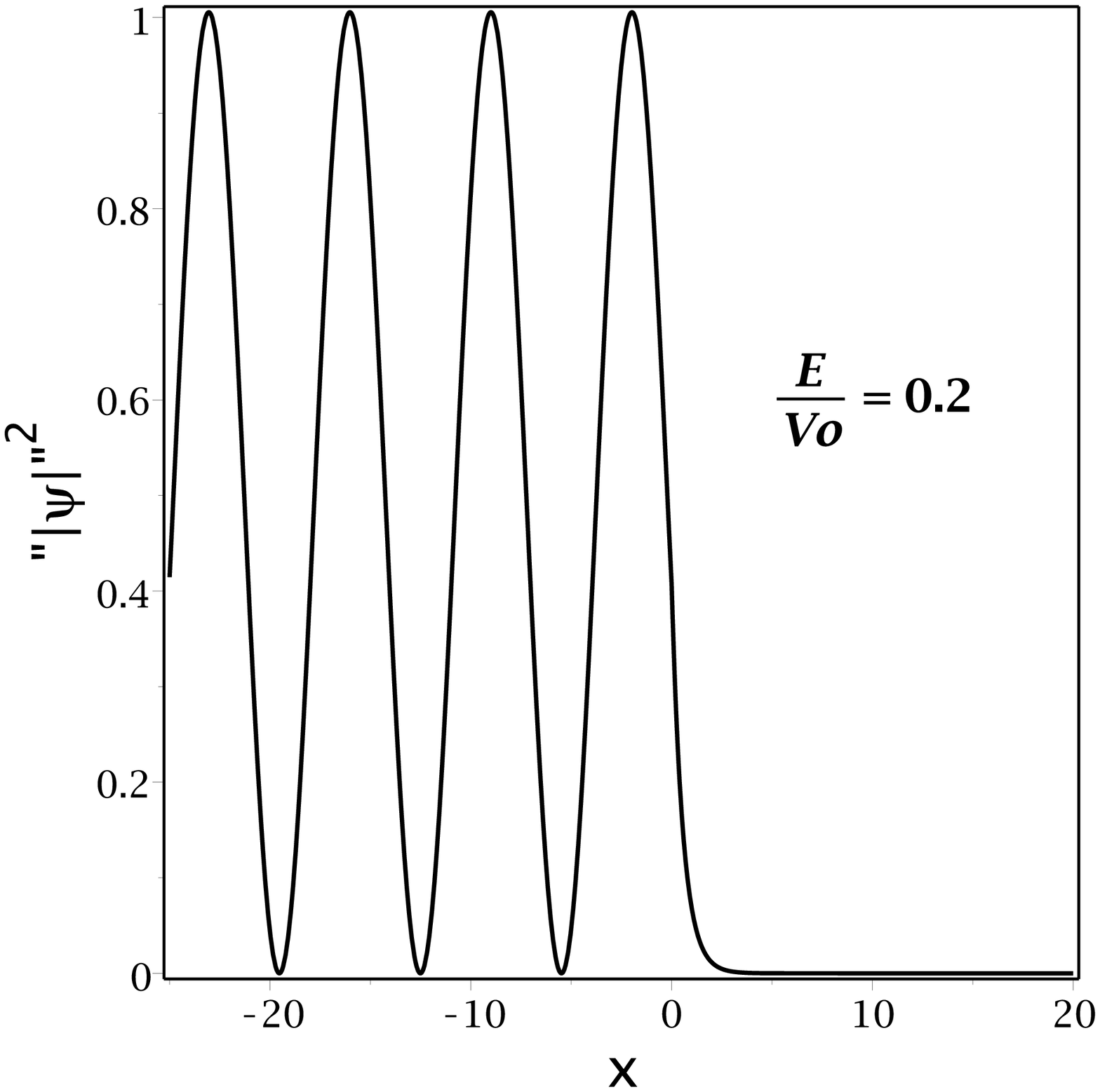}}
{\includegraphics[width=6.8cm,height=6.8cm]{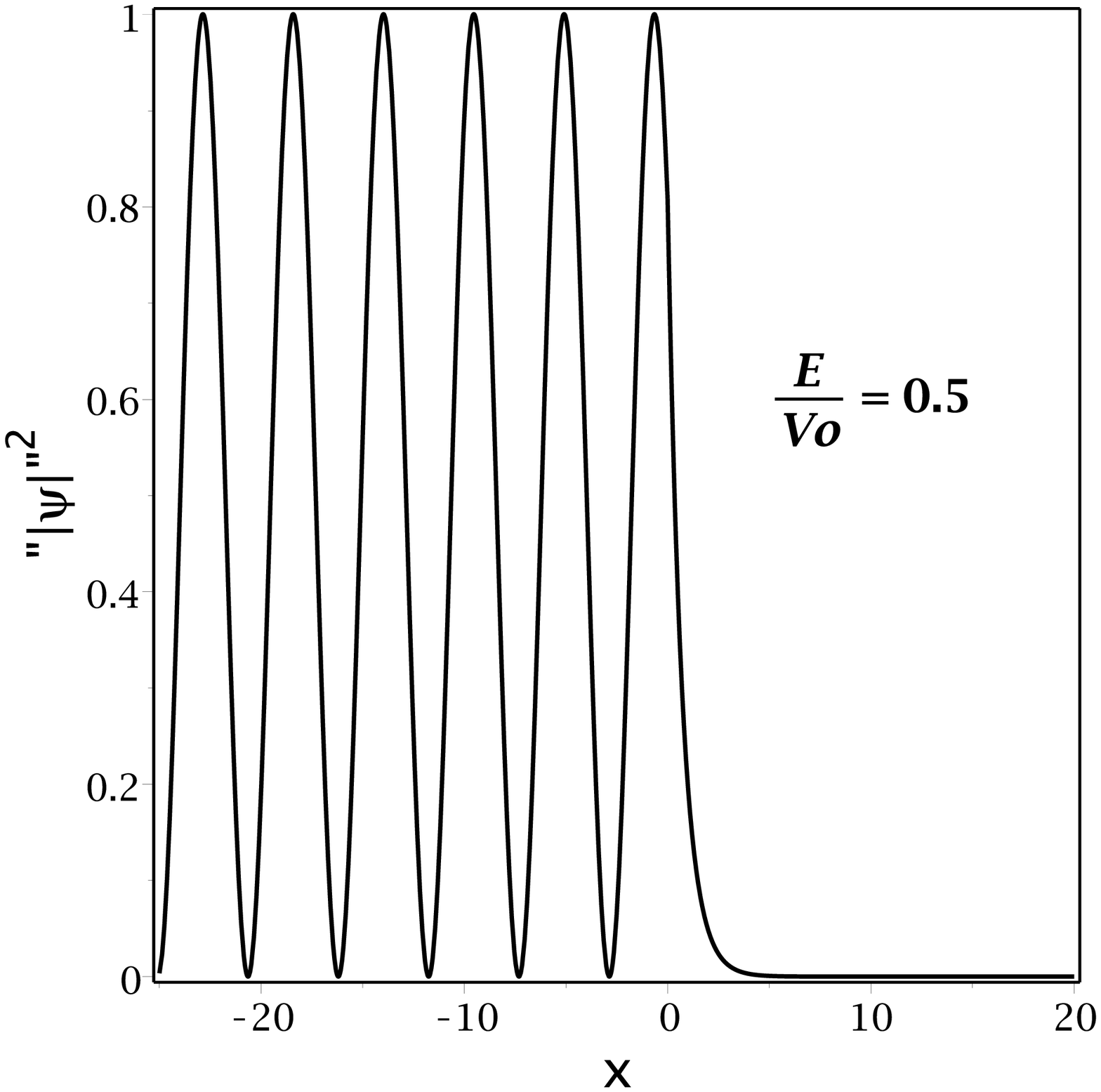}}\hspace{1cm}
{\includegraphics[width=6.8cm,height=6.8cm]{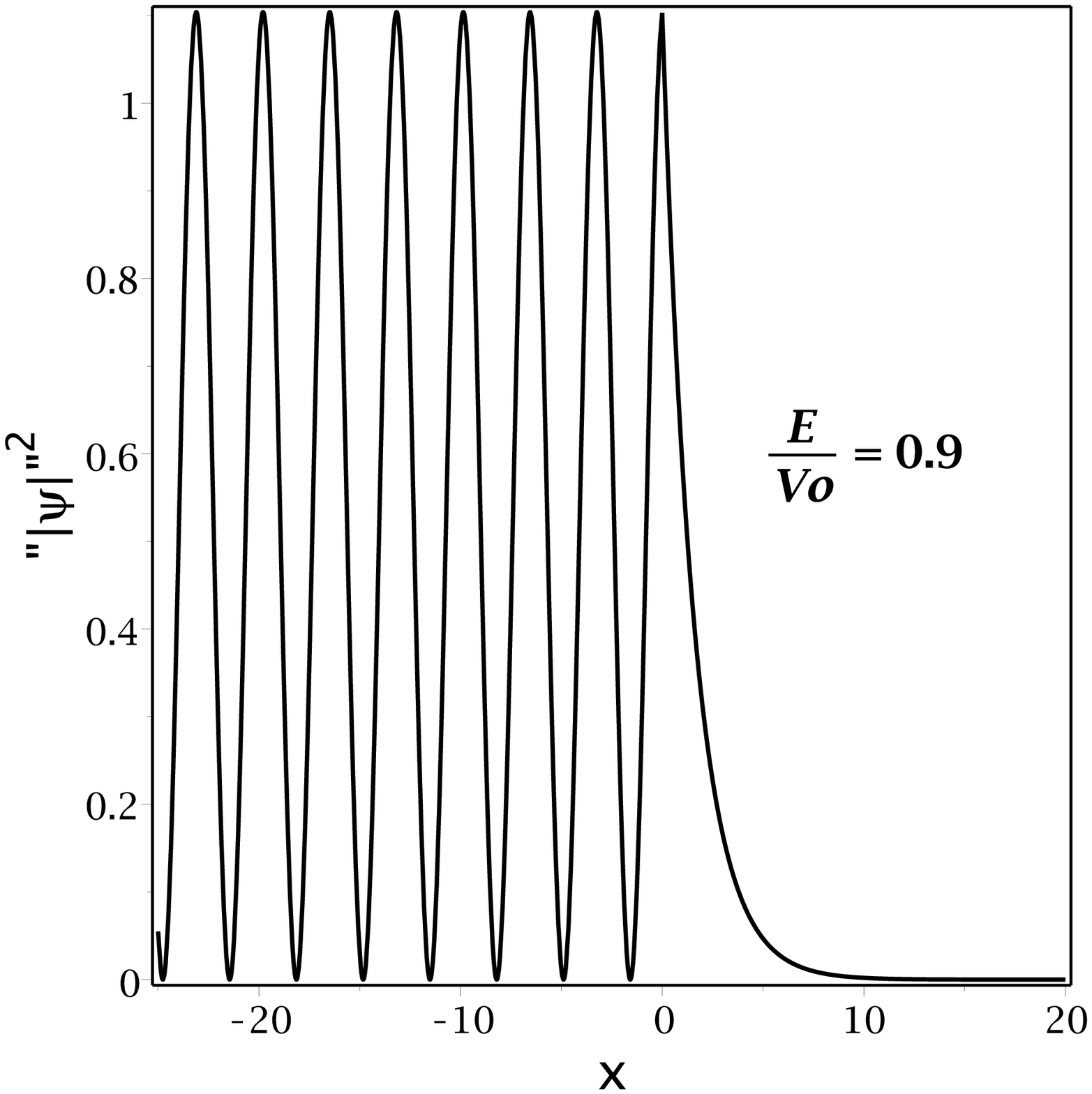}}
\caption{\label{fig_wf_E_menor_V2}M\'odulo quadrado da fun\c{c}\~ao de onda, $|\psi|^2$, para energia abaixo da altura m\'axima da barreira de potencial, $\mathcal{E}/\mathcal{V}_0=0.1,~0.2,~0.5,~0.9$, para $\delta=10.$ }
\end{figure}

\noindent\hspace{1.5cm}O segundo conjunto de gr\'aficos, Fig. (\ref{fig_wf_E_menor_V2}), representa a densidade de probabilidade para um perfil mais abrupto do potencial para $\delta = 10$ [\textit{cf}. Fig. (\ref{pot})], apresentando um comportamento esperado mais pr\'oximo de um potencial degrau t\'ipico.
\\
\section{Considera\c{c}\~oes finais \label{sec:conclusion}}

\noindent\hspace{1.5cm} Neste trabalho, discutimos detalhadamente as solu\c{c}\~oes da equa\c{c}\~ao de Schr\"odinger estacion\'aria para um problema antigo e pouco, ou quase nada, discutido nos livros did\'aticos de mec\^anica qu\^antica, a saber, o problema do potencial  deform\'avel hiperb\'olico tangente. Discutimos os limites assint\'oticos da equa\c{c}\~ao de Schr\"odinger, apresentamos as solu\c{c}\~oes anal\'iticas do problema e, a partir delas, o caso limite do potencial degrau quando o par\^ametro de deforma\c{c}\~ao cresce muito. Mostramos como obter todas as rela\c{c}\~oes relevantes para o potencial degrau abrupto a partir das rela\c{c}\~oes para as solu\c{c}\~oes do potencial hiperb\'olico.

\section{Agradecimentos}
Este trabalho \'e financiado parcialmente pelo Conselho Nacional de Desenvolvimento Cient\'ifico e Tecnol\'ogico - CNPq sob n\'umero 312251/2015-7 e pela Universidade Estadual do Cear\'a - UECE atrav\'es do Programa Institucional de Bolsas IC/UECE. 

\end{document}